\documentclass[12pt]{article}
\usepackage{graphicx}

\begin{document}

\begin{center}
\Large {\bf  Quantum Entanglement on Boundaries}
\end{center}

\bigskip
\bigskip

\begin{center}
D.V. Fursaev
\end{center}

\bigskip
\bigskip

\begin{center}
{\it Dubna International University \\
     Universitetskaya str. 19\\
     141 980, Dubna, Moscow Region, Russia\\

  and\\

  the Bogoliubov Laboratory of Theoretical Physics\\
  Joint Institute for Nuclear Research\\
  Dubna, Russia\\}
 \medskip
\end{center}

\bigskip
\bigskip

\begin{abstract}
Quantum entanglement in 3 spatial dimensions
is studied in systems with physical boundaries when an entangling surface intersects
the boundary. We show that 
there are universal logarithmic boundary terms in the entanglement R\'{e}nyi entropy and 
derive them
for different conformal field theories and geometrical configurations. 
The paper covers such topics as 
spectral geometry on manifolds with conical singularities crossing the boundaries, the dependence of the entanglement 
entropy on mutual position of the boundary and the entangling surface, effects of
acceleration and rotation of the boundary, relations of coefficients in the 
trace anomaly to coefficients in the boundary logarithmic terms in the entropy. The computations are done for scalar, spinor and gauge fields.
\end{abstract}

\newpage

\section{Introduction}\label{intr}

Studying entanglement of degrees of freedom of a quantum system on its physical boundary
is an interesting and motivated problem. It may tell us about features which distinguish 
boundary correlations from those in the bulk, explain how the structure of the boundary,
its composition, shape, roughness, and etc affect the strength of the correlations. One of
the possibilities to probe this phenomenon is to consider entanglement 
across a spatial 'entangling' surface (denoted further as $\cal B$) and to assume that
$\cal B$ ends on the boundary of the system. 
Some of the effects at the
intersection of the two surfaces may be rather complicated since 
they depend on the material the boundary is made of. 
As studies of the Casimir effect in real media
show, taking into account these properties may be a challenging problem. 
In this paper we follow a common approach and assume idealized boundaries
with conditions of the Dirichlet type.
We consider such an approach 
as a first step toward understanding more physical conditions.

We work with quantum field
theories in three spatial dimensions (a four dimensional spacetime). To quantify the entanglement we use entanglement entropy and
entanglement R\'{e}nyi entropies (ERE). In a free quantum field
theory with a spatial boundary $\partial {\cal M}$ the leading terms in the entanglement 
entropy have been studied in simple configurations: in  
\cite{Fursaev:2006ng} for flat rectangular boundaries, and in
\cite{Hertzberg:2010uv} for a system in a waveguide geometry. In both cases $\cal B$
was assumed to be orthogonal to the boundary. These results can be extended to 
R\'{e}nyi entropies which have a similar geometrical structure. By taking into account
results of \cite{Fursaev:2006ng}-\cite{Fursaev:2012mp}
one expects the following asymptotic behaviour of ERE in a four dimensional spacetime:
\begin{equation}\label{1.1}
S^{(n)}({\cal B})\simeq \frac 12 \Lambda^{2}s^{(n)}_{~2}+\Lambda s^{(n)}_{~3}
 +s^{(n)}_{~4}\ln(\Lambda\mu)+...~~,
\end{equation}
where $n$ is an order of a R\'{e}nyi entropy (see the definition in 
sec. \ref{Def}),
$\Lambda$ is an ultraviolet cutoff,
$\mu$ is a typical scale of the theory. The canonical mass dimensions of  $\Lambda$ and $\mu$ are $+1$ and $-1$, respectively. The entanglement 
entropy follows from (\ref{1.1}) in  the limit $n\to 1$. As a result of idealized
conditions microscopical parameters such as, for example, the atomic spacing of 
the boundary material, do not appear in (\ref{1.1}).

In (\ref{1.1}) the leading term  $s^{(n)}_{~2}$ is proportional to the area 
of ${\cal B}$ . Intersection of ${\cal B}$ with $\partial {\cal M}$
is a curve $\cal C$. We call $\cal C$ the 'entangling' curve. 
Computations show \cite{Fursaev:2006ng} that the next 
term $s^{(n)}_{~3}$ is proportional to the length of $\cal C$.

The focus of this paper is on boundary effects in the logarithmic term 
$s^{(n)}_{~4}$ where one expects a combination of invariants on $\cal B$ and $\cal C$.
A primary motivation is that $s^{(n)}_{~4}$ is related to the trace anomaly of the stress
tensor. Another motivation is that in lower dimensions the 
boundary contributions to analogous logarithmic 
terms carry an important physical information. As was shown in \cite{Calabrese:2009qy} 
in two-dimensional conformal theories with a
boundary \cite{Cardy:2004hm} the logarithmic term depends on the so-called 
boundary entropy or $g$-function introduced in \cite{Affleck:1991tk}.
The $g$-function decreases under the renormalization, from a critical
point to a critical point \cite{Friedan:2003yc}. One may expect that 
boundary terms in four dimensions may have similar features.

In four-dimensional conformal theories 
there is a universal part
of $s^{(n)}_{~4}$ depending on geometrical properties 
on entangling curve $\cal C$. More precisely,
\begin{equation}\label{1.2}
s^{(n)}_{~4}=a(n)F_a+c(n)F_c+b(n)F_b+d(n)F_d+e(n)F_e+z(n)~~,
\end{equation}
where functionals $F_a,F_b,F_c$ are set on $\cal B$ ($F_a$ being a topological
invariant of $\cal B$) while $F_d$, $F_e$ are defined
entirely on $\cal C$. The numbers $z(n)$ are related to zero modes.  All these 5 functionals are scale invariant and independent.
Coefficients $a(n),b(n)$ and  
$F_a,F_b,F_c$  have been determined in \cite{Fursaev:2012mp} for the case
without boundaries.
The coefficients are 3d order polynomials of $\gamma_n=1/n$.
(Note that $a(\gamma_n)$, $b(\gamma_n)$, $c(\gamma_n)$  in \cite{Fursaev:2012mp} correspond to 
$a(n)$, $b(n)$, $c(n)$ in (\ref{1.2}).)
It is the identification of $F_d$, $F_e$, and calculation 
of one of the coefficients, $d(n)$,  is the main technical problem solved in the present paper for the case of conformal theories.

The paper is organized as follows. In Sec. \ref{Def} we define the R\'{e}nyi entropy, introduce necessary elements of spectral
geometry, and 
describe a relation of $s^{(n)}_p$ in 
(\ref{1.1}) to heat kernel coefficients. Special attention is paid here to
boundary conditions for scalar, spinor and gauge fields which preserve the conformal invariance.
Section \ref{HKC} is devoted to a heat kernel
coefficient $A_4$ of a Laplace operator on a 4-dimensional manifold with conical singularities
and boundaries. $A_4$ allows one to determine $s^{(n)}_4$. Conformal properties of 
the Laplace operator and conformal invariance
of boundary conditions
are assumed. When a co-dimension 2 surface where conical singularities are located 
(an entangling surface $\cal B$) crosses the boundary, $A_4$ acquires boundary terms
located on a curve ${\cal C}={\cal B}\cap \partial {\cal M}$.
If the boundary is smooth we prove that only two independent terms, $F_d$ and $F_e$,
on $\cal C$
can exist in $A_4$ and we fix their structure by the conformal invariance. 
We then compute a coefficient function $d(n)$ at $F_d$. The section ends with 
examples of $F_d$ and $F_e$ for some geometrical configurations in a flat spacetime.
In Sec. \ref{Phys} we discuss a number of physical applications where
the boundary entanglement plays a key role. We consider a difference of two entanglement entropies $\Delta S$
for a fixed model and the same boundary conditions. We assume that entangling surfaces and curves for the two systems have, respectively, equal areas and lengths. The two entropies then differ by the 
logarithmic terms presented in (\ref{1.2}). In a flat spacetime $\Delta S$ may depend
only on the entanglement on boundaries.  The results of this section include: a study
of scaling properties of $\Delta S$ for different geometrical configurations and 
for boundaries which have non-zero local rotation and acceleration. 
The acceleration and rotation effects follow from the structure of $F_d$. 
In a simplest case $F_d$ may be
proportional to an integral over $\cal C$ of the acceleration in the direction orthogonal 
to $\partial {\cal M}$. A discussion of the above results with an emphasis on open problems is given in sec. \ref{concl}. In particular, we analyse possible relations of coefficients in the trace anomaly to a coefficient $d$ at $F_d$ in the entropy and discuss its relevance 
in studying renormalization group flow in the boundary terms.
Appendix A collects all geometrical notations. Unavoidable technicalities are left for Appendix B (calculations of  the heat coefficients for different spins to fix $d(n)$)
and Appendix C (a 3+1 decomposition of the boundary extrinsic curvature 
tensor in a Killing 
frame of reference).

\begin{figure}[h]
\begin{center}
\includegraphics[height=8cm,width=11cm]{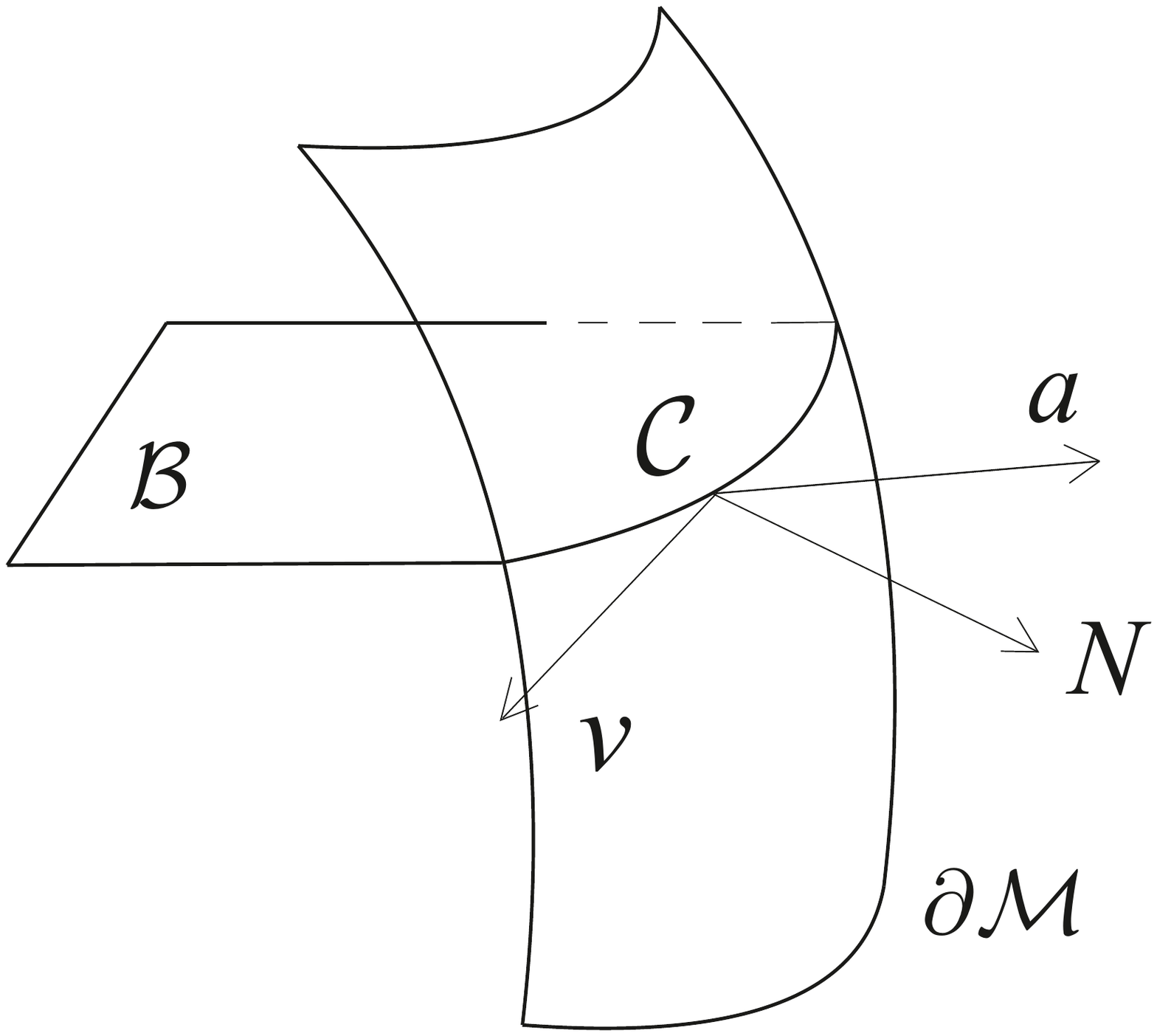}
\caption{\small{Entangling surface
$\cal B$ crossing the boundary $\partial {\cal M}$, $N$ is a normal vector to 
$\partial {\cal M}$, $v$ is a tangent vector to entangling curve $\cal C$, $a$ is the acceleration vector of $\cal C$.}}
\label{F1}
\end{center}
\end{figure}

\section{Definitions}\label{Def}
\setcounter{equation}0
\subsection{Entropy}

We consider a quantum field theory on a static four-dimensional spacetime with constant
time sections $\Sigma$.
The state of the system is specified by a density matrix  $\hat{\rho}$ and 
the entanglement between 
spatially separated parts, $A$ and $B$, with a common boundary $\cal B$ is determined
by a reduced density matrix. Say, for the region $A$, the reduced matrix is
defined as
\begin{equation}\label{2.1}
\hat{{\rho}}_A=\mbox{Tr}_B \hat{\rho}~~,
\end{equation}
by taking trace over the states located in the region $B$.
The entanglement R\'{e}nyi entropy of an order $n$ is defined as
\begin{equation}\label{2.2}
S^{(n)}_A={\ln \mbox{Tr}_A \hat{{\rho}}^{~n}_A \over 1-n}~~,
\end{equation}
where $n$ is a non-negative parameter, $n\neq 1$. 
A related notion, the corresponding entanglement entropy
\begin{equation}\label{2.3}
S_A=-\mbox{Tr}_A \hat{{\rho}}_A\ln \hat{{\rho}}_A~~,
\end{equation}
follows from (\ref{2.2}) in the limit $n\to 1$.
In the rest part of this paper we consider integer values $n=2,3,...$.
In local terms $s^{(n)}_p$ the limit $n\to 1$ does not pose a problem.

Analogously, one can define the R\'{e}nyi entropy $S^{(n)}_B$ for a density matrix
obtained by integrating over states in region $A$. There is an important symmetry property,
$S^{(n)}_B=S^{(n)}_A$, when the system is in a pure state. In what follows we do not write
explicitly the indexes $A$ or $B$ in the 
entropy. Properties of the entropy we discuss do not depend on the choice of the reduction procedure even when the state is not pure.

For technical reasons it is convenient to choose thermal density matrix,
$\hat{\rho}=e^{-\hat{H}/T}/Z(T)$, where $T$ is the temperature, $\hat{H}$ is a Hamiltonian,
$Z(T)=\mbox{Tr}~\exp(-\hat{H}/T)$ is a partition function. One recovers the vacuum state
in the limit $T\to 0$. The finite-temperature theory corresponds to a Euclidean
four-dimensional manifold $\cal M$ with constant time sections $\Sigma$. 
The orbits of the Killing vector field generating translations in Euclidean time
are the circles $S^1$ with the length equal $1/T$.

The entropy 
can be written as 
\begin{equation}\label{2.4}
S^{(n)}(T)={1\over 1-n}\left(\ln Z(n,T)-n \ln Z(T)\right)~~,
\end{equation}
\begin{equation}\label{2.5}
Z(n,T)=\mbox{Tr}_A\left(\mbox{Tr}_B~e^{-\hat{H}/T}\right)^n~~,
\end{equation}
Here $Z(n,T)$ is an 'entanglement partition function', $Z(1,T)=Z(T)$. 

In a quantum field theory the partition function $Z(T)$
is represented as a functional integral over field configurations
which live on ${\cal M}$. Analogously, $Z(n,T)$ can be written in terms of a path integral
where field configurations
are set on a 'replicated' manifold ${\cal M}_n$ which is glued from $n$ copies (replicas) of $\cal M$
along some cuts which meet on $\cal B$, see \cite{Fursaev:2006ng}. 
${\cal M}_n$
are locally identical to $\cal M$ but have conical singularities
on $\cal B$ with the length of a small unit circle around each point on $\cal B$ 
equals $2\pi n$. By the definition ${\cal M}_1={\cal M}$.

For a free QFT the partition function $Z(n,T)$ is defined in terms of a regularized determinant $(\det\Delta)^{\mp 1/2}$ of a Laplace operator $\Delta$. 
The base manifold for the
Laplace operators is ${\cal M}_n$.  
The details of these constructions are described, e.g.
in \cite{Fursaev:2012mp}. 

\subsection{Heat kernels}

Let $A_p(\Delta)$ be heat coefficients for the asymptotic expansion of the heat kernel of a Laplacian $\Delta$, 
\begin{equation}\label{2.6}
K(\Delta;t)=\mbox{Tr}~e^{-t\Delta}\simeq \sum_{p=0} A_p(\Delta)~t^{(p-4)/2}~~,~~t\to 0~~.
\end{equation}
The number of spacetime dimensions in (\ref{2.6}) is 4.
By taking into account (\ref{2.4}), (\ref{2.5}) one can show \cite{Fursaev:2012mp} that 
entropies in (\ref{1.1}) are expressed as
\begin{equation}\label{2.7}
s^{(n)}_{~p} =\eta {nA_p(1)-A_p(n) \over n-1}~~,~~p\neq 4~~,
\end{equation}
\begin{equation}\label{2.8}
s^{(n)}_{~4} =\eta{nA_4(1)-A_4(n) \over n-1}+z(n)~~,
\end{equation}
where $A_p(n)$ are heat coefficients $A_p(\Delta)$ for $\Delta$ 
on ${\cal M}_n$, $\eta=+1$ for Bosons and $\eta=-1$ for Fermions. If $\Delta$ 
on ${\cal M}_n$ has a non-vanishing number of zero
modes $N_{\mbox{\tiny{zm}}}(n)$, the zero modes have to be
excluded from $\det\Delta$.
This subtraction affects only the coefficient $A_4(\Delta)$ and it yields the
last term in the r.h.s.
of $s^{(n)}_{~4}$ in (\ref{2.8}). One can show that
\begin{equation}\label{2.9}
z(n)=-
\eta{nN_{\mbox{\tiny{zm}}}(1)-N_{\mbox{\tiny{zm}}}(n) \over n-1}~~.
\end{equation}
For given boundary conditions the number of zero modes and $z(n)$ are determined
only by topologies of the background manifolds. Contribution of $z(n)$ in (\ref{2.8})
can be important. For example, in a pure 2D gauge theory without this contribution
the entanglement entropy computed by the method of conical singularities would be non-trivial, see \cite{Fursaev:1996uz}.

We consider manifolds whose constant time sections $\Sigma$ have boundaries
$\partial \Sigma$. Boundary $\partial {\cal M}$ of the 4-dimensional manifold 
is $\partial {\cal M}\sim S^1 \times \partial \Sigma$. 

The definitions related to the geometry of entangling surface $\cal B$
are as follows (see Appendix \ref{C-notations}). We assume that $\cal B$ crosses $\partial {\cal M}$.
Since  $\cal B$  lies in a constant time section $\Sigma$ 
it also crosses $\partial \Sigma$. The intersection
of $\cal B$ and $\partial \Sigma$ (or $\partial {\cal M}$) is denoted 
by $\cal C$ and is called the entangling curve, see Fig. \ref{F1}. We also introduce 
different unit vectors: $v$ is a tangent vector to $\cal C$, $a=\nabla_v v$ is an acceleration of $\cal C$,
$N$ is an outward pointing
normal vector to $\partial {\cal M}$, $p_i$ is an ortho-normalized pair of normal vectors to $\cal B$, 
and $m_a$ is an ortho-normalized pair of vectors at $\cal C$ which are orthogonal
to $v$ and $N$. We denote $R$, $R_{\mu\nu}$,
$R_{\mu\nu\lambda\rho}$ the scalar curvature, the Ricci tensor and the Riemann tensor of the regular part of ${\cal M}_n$, respectively.

\subsection{Operators and boundary conditions}\label{BC}

In this paper we consider only models with massless conformal scalar and spinor fields,
as well as gauge models. We choose boundary  conditions which are invariant 
under conformal transformations  defined in Sec. \ref{HKC}. The scalar Laplacian is 
taken as $\Delta^{(0)}=-\nabla^2+\frac 16 R$, and the boundary condition
is the Dirichlet condition
\begin{equation}\label{2.10-d}
\varphi\mid_{\partial {\cal M}}=0~~,
\end{equation}
which is manifestly conformally invariant.

Quantization of an Abelian gauge field $V_\mu$ is considered in the Lorentz gauge 
$\nabla V=0$. The corresponding vector Laplacian is
$(\Delta^{(1)})^\nu_\mu=-\nabla^2\delta^\nu_\mu+R^\nu_\mu$ and the 
Laplacian for ghosts is $\Delta^{(\mbox{\tiny gh})}=-\nabla^2$.
We use the following boundary condition:
\begin{equation}\label{2.10-g}
N^\mu F_{\mu\nu}\mid_{\partial {\cal M}}=0~~,
\end{equation}
where $F_{\mu\nu}=\nabla_\mu V_\nu-\nabla_\nu V_\mu$.
This condition is manifestly gauge and conformally 
invariant. It requires that components of an electric field which normal to $\partial {\cal M}$
and components of the magnetic field which tangential to $\partial {\cal M}$
vanish on the boundary. 
The condition like this is physically motivated when the boundary is a perfect 
conductor. In the Lorentz gauge we use the so called absolute boundary conditions
\cite{Vassilevich:2003xt} 
\begin{equation}\label{2.10-gv}
V_N\mid_{\partial {\cal M}}=0~~,~~(N^\mu\nabla_\mu 
V^\nu_{\parallel}+K_{\mu}^\nu V^\mu_{\parallel})\mid_{\partial {\cal M}}=0
\end{equation}
where $K_{\mu\nu}$ is an extrinsic curvature tensor 
of $\partial {\cal M}$,
$V_N=N^\mu V_\mu$ and $V_{\parallel}$ are, respectively, normal and tangential components of the vector field to $\partial {\cal M}$.
The corresponding boundary condition for a ghost field $c$ is
\begin{equation}\label{2.10-gg}
\partial_N c\mid_{\partial {\cal M}}=0~~.
\end{equation}
Physical condition 
(\ref{2.10-g}) follows from (\ref{2.10-gv}), (\ref{2.10-gg}). 

In case of a massless Dirac field $\psi$ the operator is
$\Delta^{(1/2)}=(i\gamma^\mu\nabla_\mu)^2$ and we require that
\begin{equation}\label{2.10-dd}
\Pi_- \psi\mid_{\partial {\cal M}}=0~~,
\end{equation}
where $\Pi_-=\frac 12 (1\pm i\gamma_\ast N^\mu \gamma_\mu)$, and 
$\gamma_\ast$ is a chirality gamma matrix. The physical meaning of (\ref{2.10-dd})
is that the normal component of the spinor current vanishes on the
boundary. Condition (\ref{2.10-dd}) does not break conformal invariance.

If a 4D classical theory is conformally invariant 
one can show that the heat coefficient $A_4(\Delta)$ for the corresponding Laplacian
$\Delta$ remains invariant under local conformal 
transformations, see e.g. \cite{Dowker:1989gw}, \cite{Fursaev:2011zz}. In general, gauge fixing 
procedure breaks the conformal invariance in gauge models (also on the level of boundary conditions (\ref{2.10-gv}),(\ref{2.10-gg})). For 
the gauge field in the Lorentz gauge we always consider 
a 'total' heat coefficient
\begin{equation}\label{2.11}
A_4^{(\mbox{\tiny gauge})}=A_4(\Delta^{(1)})-2A_4(\Delta^{(\mbox{\tiny gh})})~~
\end{equation}
since this is a gauge invariant combination which determines the one-loop divergences in the effective action (in the dimensional regularization, for example). 
A proof that these divergences are conformally invariant 
can be found in \cite{Buchbinder:1984}. An explicit demonstration 
of conformal invariance of the part of $A_4^{(\mbox{\tiny gauge})}$ 
which appears due to conical singularities in the absence of boundaries is presented in \cite{Fursaev:2012mp}.

\section{Spectral geometry}\label{HKC}
\setcounter{equation}0

\subsection{Heat kernels and conical singularities located on boundaries}

Equations (\ref{2.7}), (\ref{2.8}) show that leading terms in 
the entanglement entropies are related to contributions
from conical singularities to corresponding heat kernel coefficients. 
In this section, therefore, we study 
the spectral geometry on manifolds which, like ${\cal M}_n$, 
have conical singularities located on a co-dimension 2 
hypersurface $\cal B$ which crosses the boundary. Such manifolds 
close to the intersection of $\cal B$ and $\partial {\cal M}_n$
have structure 
$C_n\times {\cal C}$. A conical angle 
of the conical space $C_n$ is $2\pi n$. Boundary conical singularities produce extra contributions
to heat coefficients in a form of local invariant functionals given on $\cal C$. Our aim
is to fix the structure of these functionals.

The 
dimensionality $A_p(\Delta)$ is $L^{d-p}$, where $d$ is the number of spacetime 
dimensions and $L$ is a length parameter. Therefore, functionals on $\cal C$ must
be integrals of curvature invariants which have the dimensionality 
$L^{d-p-(d-3)}=L^{3-p}$. Since dimensionality of curvature invariants should be non-positive
we expect that relevant boundary terms appear in $A_p(\Delta)$ only if $p\geq 3$. This is 
exactly what one can learn from particular geometries \cite{Fursaev:2006ng},\cite{Hertzberg:2010uv}. The analysis of \cite{Fursaev:2006ng} shows that 
boundary conical singularities yield a contribution to $A_3(\Delta)$
(proportional to volume of $\cal C$) which is not universal
and depends on boundary conditions. According to (\ref{2.7})
the contribution $s^{(n)}_{~3}$
is not universal as well.

In the rest of this paper we focus only on boundary terms in $A_4(\Delta)$. One 
may write $A_4(\Delta)=A_4(n)=nA_4(n=1)+\bar{A}_4(n)$ where $nA_4(n=1)$ is a part 
of the heat coefficient  determined on a regular domain of ${\cal M}_n$ in a standard way. The part of the heat coefficient which depends on conical singularities 
can be written as
\begin{equation}\label{3.1}
\bar{A}_4(n)=
\bar{a}(n)F_a+\bar{c}(n)F_c+\bar{b}(n)F_b+\bar{d}(n)F_d+\bar{e}(n)F_e~~,
\end{equation}
and if one takes into account (\ref{1.2}) and (\ref{2.8}),
\begin{equation}\label{3.1b}
\bar{A}_4(n)=
\eta(1-n)(s^{(n)}_{~4}-z(n)) 
\end{equation}
$$
\bar{a}(n)=\eta(1-n)a(n)~~,~~\bar{b}(n)=\eta(1-n)b(n)~~,~~\bar{c}(n)=\eta(1-n)c(n)~~,
$$
\begin{equation}\label{3.1c}
\bar{d}(n)=\eta(1-n)d(n)~~,~~\bar{e}(n)=\eta(1-n)e(n)~~.
\end{equation}
Since we require that theory is conformally invariant,
quantities $F_a$ - $F_e$ are defined as a number of independent
conformal invariants. By the definition,
$F_d$, $F_e$ are present
only for boundary conical singularities.
We call them boundary terms.
In the absence of boundary conical singularities
invariant functionals, $F_a$, $F_b$, and $F_c$, have been determined  
in \cite{Fursaev:2012mp} on the base of previous results, see 
references therein.

The quantity $F_a$ has been expressed in \cite{Fursaev:2012mp} in terms
of
the Euler characteristic $\chi_2$ of ${\cal B}$. It is convenient to keep this
relation also in case of boundary conical singularities
and define 
\begin{equation}\label{3.2}
F_a=-2\chi_2[{\cal B}]=-{1 \over 2\pi}\left[
\int_{{\cal B}}\sqrt{\sigma}d^2x~R({\cal B})+2\int_{\cal C}k_B ds\right]~~
\end{equation}
by adding a standard boundary term.
Here $R({\cal B})$ is the scalar curvature of $\cal B$ and $k_B$ is 
an extrinsic curvature of $\cal C$ in $\cal B$, $ds$ is the arclength 
of $\cal C$. The functional $F_c$ is defined as
\begin{equation}\label{3.3}
F_c={1 \over 2\pi}\int_{{\cal B}}\sqrt{\sigma}d^2x~C_{ijij}~~,
\end{equation}
\begin{equation}\label{3.4}
C_{ijij}=C_{\mu\nu\lambda\rho}p_i^\mu p_j^\nu p_i^\lambda p_i^\rho~~.
\end{equation}
$p_i$, $i=1,2$, are two unit mutually orthogonal normal vectors 
to ${\cal B}$. Summation over repeated indexes is implied.
The Weyl tensor in four dimensions is
$$
C_{\mu\nu\lambda\rho}=R_{\mu\nu\lambda\rho}+{1 \over 2}\left(g_{\mu\rho}R_{\nu\lambda}
+g_{\nu\lambda}R_{\mu\rho}-g_{\mu\lambda}R_{\nu\rho}-g_{\nu\rho}R_{\mu\lambda}\right)
$$
\begin{equation}\label{3.5}
+\frac 16 R\left(g_{\mu\lambda}g_{\nu\rho}-g_{\mu\rho}g_{\nu\lambda}\right)~~.
\end{equation}
Finally,
\begin{equation}\label{3.6}
F_b={1 \over 2\pi}\int_{{\cal B}}\sqrt{\sigma}d^2x~\left({1 \over 2} k_i^2-\mbox{Tr}(k_i^2)\right)=-{1 \over 2\pi}\int_{{\cal B}}\sqrt{\sigma}d^2x \mbox{Tr}(\hat{k}_i^2)~~~,
\end{equation}
where $(k_i)_{\mu\nu}=P_{\mu}^\lambda P_{\nu}^\rho (p_i)_{\lambda;\rho}$ 
are extrinsic curvatures of $\cal B$, 
$P_{\mu}^\lambda=\delta_{\mu}^\lambda-(p_i)_{\mu}(p_i)^\lambda$ 
is a projector on 
directions tangent to $\cal B$,
$k_i=g^{\mu\nu}(k_i)_{\mu\nu}$, $\mbox{Tr}(k_i^2)=(k_i)_{\mu\nu}(k_i)^{\mu\nu}$. The traceless part $(\hat{k}_i)_{\mu\nu}=
(k_i)_{\mu\nu}-\frac 12 g_{\mu\nu}k_i$  changes homogeneously
under scaling transformations. 

Let us emphasize that $F_a$, $F_b$, and $F_c$ are universal functionals which do not 
depend on boundary conditions. Explicit expressions
of coefficients $\bar{a}(n)$,  $\bar{c}(n)$ for different CFT's 
can be found in \cite{Fursaev:2012mp}.

\subsection{Conformal invariants on entangling curve}

Let us discuss now boundary functionals in (\ref{3.1}). They have  
a form of integrals over $\cal C$ of some curvature invariants.
Since $A_4(\Delta)$ is dimensionless 
in $d=4$ the curvature invariants should have dimensionality $L^{-1}$.
The only appropriate material the invariants can be made of are 
different extrinsic curvatures. There may be three sorts of terms related to 
$\partial {\cal M}$, $\cal B$, and $\cal C$. 

\bigskip

{\bf Curvatures of the spacetime boundary}. 
We define the extrinsic curvature tensor of $\partial {\cal M}$ as
$K_{\mu\nu}=H_{\mu}^\lambda H_{\nu}^\rho N_{\lambda;\rho}$, where $H_{\mu}^\nu=
\delta_{\mu}^\nu-N_{\mu}N^\nu$. According to definitions set in Sec. \ref{Def} one has the following
invariant quantities on $\cal C$: $K_{ab}=K_{\mu\nu}m_a^\mu m_b^\nu$, 
$K_{va}=K_{\mu\nu}v^\mu m_a^\nu$, $K_{vv}=K_{\mu\nu}v^\mu v^\nu$, where
$v$ is a tangent vector to $\cal C$
and $m_a$ is an ortho-normalized pair of vectors at $\cal C$ which are orthogonal
to $v$ and $N$. 
Since the boundary terms should not depend on the choice of the basis they must obey
an additional $O(2)$ symmetry related to a rotation of the basis $m_1,m_2$. This requirement
leaves only two possible terms: $K_{ab}\delta^{ab}$ and $K_{vv}$, or, as an equivalent option,
$K$ and $K_{vv}$, where $K=K_{\mu}^{\mu}$. This leaves us with the following
functional:
\begin{equation}\label{3.7}
F_d=
-{1 \over 2\pi}\int_{{\cal C}}ds~(K-3K_{vv})={3 \over 2\pi}\int_{{\cal C}}ds~\hat{K}_{vv}~~,
\end{equation}
\begin{equation}\label{c.12}
\hat{K}_{\mu\nu}=K_{\mu\nu}-\frac 13 H_{\mu\nu}K~~,
\end{equation}
where numerical coefficients are chosen just for further convenience. Functional (\ref{3.7})
is invariant under conformal transformations
\begin{equation}\label{3.8}
\bar{g}_{\mu\nu}(x)=e^{-2\omega(x)}g_{\mu\nu}(x)~~,
\end{equation}
\begin{equation}\label{3.9}
\bar{K}_{\mu\nu}=e^{-\omega}\left[K_{\mu\nu}-H_{\mu\nu}~\omega_{,N}\right]~~,
~~\bar{N}^\mu=e^{\omega}N^\mu~~,
~~\bar{v}^\mu=e^{\omega}v^\mu~~,
\end{equation}
where $\omega_{,N}=N^\lambda\omega_{,\lambda}$. The traceless part (\ref{c.12}) transforms homogeneously.

\bigskip

{\bf Extrinsic curvatures of entangling surface on the boundary}.
Consider the  extrinsic curvature tensors
$(k_i)_{\mu\nu}$ of $\cal B$. Since they are 2 by 2 symmetric matrices
we can use only three sorts of coordinate invariants on $\cal C$: 
$(k_i)_{vv}=(k_i)_{\mu\nu}v^\mu v^\nu$, $(k_i)_{vl}=(k_i)_{\mu\nu}v^\mu l^\nu$, 
$(k_i)_{ll}=(k_i)_{\mu\nu}l^\mu l^\nu$, where $l$ is some unit vector which is
in a tangent space to $\cal B$ and is orthogonal to $\cal C$. We choose $l$ as outward 
directed vector. 
The term $(k_i)_{vl}$ should
be excluded since it depends on the direction of tangent vector $v$. Instead of two remaining 
quantities it is convenient to choose $k_i$ and $(k_i)_{vv}$. Next one must ensure independence 
on the choice of the pair $p_i$ and require the corresponding $O(2)$ symmetry. This can be done by 
taking $O(2)$ invariant combinations $(N\cdot p_i)k_i$ and $(N\cdot p_i)(k_i)_{vv}$.
One cannot use prefactors such as $(m_a\cdot p_i)$ since they would depend on the choice of another
pair of normal vectors $m_a$. Other vectors associated to orientation of $\cal C$ are orthogonal to
$p_i$. Thus, we come to the following functional:
\begin{equation}\label{3.10}
F_e={1 \over \pi}\int_{{\cal C}}ds~(N\cdot p_i)(\hat{k}_i)_{vv}~~,
\end{equation}
where summation over the index $i$ is implied.
Conformal invariance of (\ref{3.10}) results from (\ref{3.8}) and transformations
\begin{equation}\label{3.11}
(\bar{k}_i)_{\mu\nu}=e^{-\omega}\left[(k_i)_{\mu\nu}-h_{\mu\nu}~\omega_{,i}\right]~~,
~~\bar{p}_i^\mu=e^{\omega}p_i^\mu~~,
\end{equation}
where $\omega_{,i}=p^\lambda_i\omega_{,\lambda}$. 
The traceless part $(\hat{k}_i)_{\mu\nu}=
(k_i)_{\mu\nu}-\frac 12 g_{\mu\nu}k_i$  changes homogeneously
under (\ref{3.11}). 

Since $F_e$ depends on the traceless part 
$(\hat{k}_i)_{\mu\nu}$ of extrinsic curvatures of $\cal B$ it is
a complete boundary analogue of the bulk functional $F_b$, see (\ref{3.6}).

\bigskip

{\bf Invariants related to the entangling curve}. The curve $\cal C$ is characterized by an acceleration vector $a^\mu=\nabla_v v^\mu$. The norm
of this vector is called the curvature of $\cal C$. Scalar products of the acceleration vector with vectors orthogonal to $v$ yield other invariant structures
on $\cal C$. One can use the definition of $a$ and orthogonality 
of $N, p_i, l$ to $v$ to see that all these quantities are reduced to 
\begin{equation}\label{3.12}
(a \cdot N)=-K_{vv}~~,~~(a \cdot p_i)=-(k_i)_{vv}~~,~~(a \cdot l)=-k_B~~.
\end{equation}
Here $k_B$ is an extrinsic curvature of $\cal C$ in $\cal B$, see (\ref{3.2}). $k_B$ is not an 
independent quantity since $l$ is a linear combination of $n$ and $p_i$. Vectors $N$ and $p_i$ are linear independent in general,
and one cannot construct a conformal invariant by using 
quantities (\ref{3.12}) alone. This brings us back to invariants considered before.

We proved, therefore, that the boundary terms due to conical singularities in heat coefficient
$A_4(\Delta)$ in conformally invariant theories are given by two invariant
functionals  (\ref{3.7}), (\ref{3.10}).

\begin{table}
\renewcommand{\baselinestretch}{2}
\medskip
\caption{Coefficient functions. The used notations are $\gamma\equiv 1/n$, $d=d(1)$.}
\bigskip
\begin{centerline}
{\small
\begin{tabular}{|c|c|c|c|c|c|}
\hline
$\mbox{field}$  & $\bar{a}(n)$ & $\bar{g}(n)$ & $\bar{d}(n)$  & $d(n)$ & $d$   \\
\hline
$\mbox{real scalar}$ &  ${\gamma^4-1 \over 1440 \gamma}$
&  ${\gamma^2-1 \over 144 \gamma}$ & ${\gamma^4+10\gamma^2- 11 \over 1440\gamma}$ 
 &  ${\gamma^3+\gamma^2+11\gamma+11 \over 1440}$   & ${1 \over 60}$ \\
\hline
$\mbox{Dirac spinor}$ &  $-{7\gamma^4+30\gamma^2-37 \over 2880 \gamma}$ 
&  ${\gamma^2-1 \over 144 \gamma}$
 &  $-{7\gamma^4+10\gamma^2-17 \over 2880 \gamma}$ &
 ${7\gamma^3+7\gamma^2+17\gamma+17 \over 2880}$ & ${1 \over 60}$  \\
\hline
$\mbox{gauge Boson}$ &  ${\gamma^4+30\gamma^2+60\gamma-91 \over 720 \gamma}$
&  $-{\gamma^2+3\gamma-4 \over 36 \gamma}$
&  ${\gamma^4+10\gamma^2-11 \over 720 \gamma}$ & 
${\gamma^3+\gamma^2+11\gamma+11 \over 720}$ & ${1 \over 30}$  \\
\hline
\end{tabular}}
\bigskip
\renewcommand{\baselinestretch}{1}
\end{centerline}
\label{t1}
\end{table}

\subsection{Fixing coefficients}

We now fix coefficient $\bar{d}(n)$ by studying
particular cases.  Consider a flat spacetime and suppose that a quantum system
is in a domain $\Sigma$ with a cylinder-like boundary $\partial \Sigma$
stretched along an axis parametrized, say, by a $z$ coordinate. 
We assume also a translational invariance of $\partial \Sigma$
along the $z$ coordinate. As is shown in Appendix \ref{A-HKC} the heat coefficients 
on ${\cal M}_n$ can be computed for this problem if the entangling surface $\cal B$ 
is flat and orthogonal to $\partial \Sigma$. 
In fact, the only contribution from
conical singularities in $A_4$ comes from the boundary,  
\begin{equation}\label{3.13}
\bar{A}_4(n)=
{\bar{g}(n) \over \pi}\int_{\cal C}k_B ds~~.
\end{equation}
The coefficient functions $\bar{g}(n)$ for different spins are given in Table \ref{t1}
for boundary conditions (\ref{2.10-d}),(\ref{2.10-gv})-(\ref{2.10-dd}).
This result
can be compared with a general formula (\ref{3.1}). Definitions 
(\ref{3.2}),(\ref{3.3}),(\ref{3.6}),(\ref{3.7}),(\ref{3.10}) yield
\begin{equation}\label{3.14}
F_a=-F_d=-{1 \over \pi}\int_{\cal C}k_B ds~,~~F_b=F_c=F_e=0~~.
\end{equation}
Here we took into account that $N=l$, and, therefore, 
$K=K_{vv}=k_B$ in (\ref{3.7}), see (\ref{3.12}).
By comparing (\ref{3.13}) with (\ref{3.1}) one finds that
\begin{equation}\label{3.15}
\bar{d}(n)=\bar{a}(n)+\bar{g}(n)~~.
\end{equation}

Since $\bar{d}(n)$ are non-trivial and  $\bar{d}(n)$  and $\bar{g}(n)$ are independent 
functions the boundary functionals $F_d$ must appear in $A_4$. Results  for
$\bar{a}(n)$, $\bar{d}(n)$ are presented in Table \ref{t1}. 
This table also includes
the values of coefficient functions $d(n)$ which appear in the R\'{e}nyi entropy, Eq.
(\ref{1.2}). $d(n)$ are derived from $\bar{d}(n)$ with the help of (\ref{3.1c}).
Note that for all spins $d(n)$ allow analytical continuation to arguments $n=1$. Values
$d=d(1)$ are coefficients which are present in corresponding logarithmic terms
in entanglement entropies (\ref{2.3}). 

We do not suggest here a method how to fix  coefficient $\bar{e}(n)$. It remains as unknown
as the analogous bulk coefficient $\bar{b}(n)$.

\begin{figure}[h]
\begin{center}
\includegraphics[height=9.5cm,width=12cm]{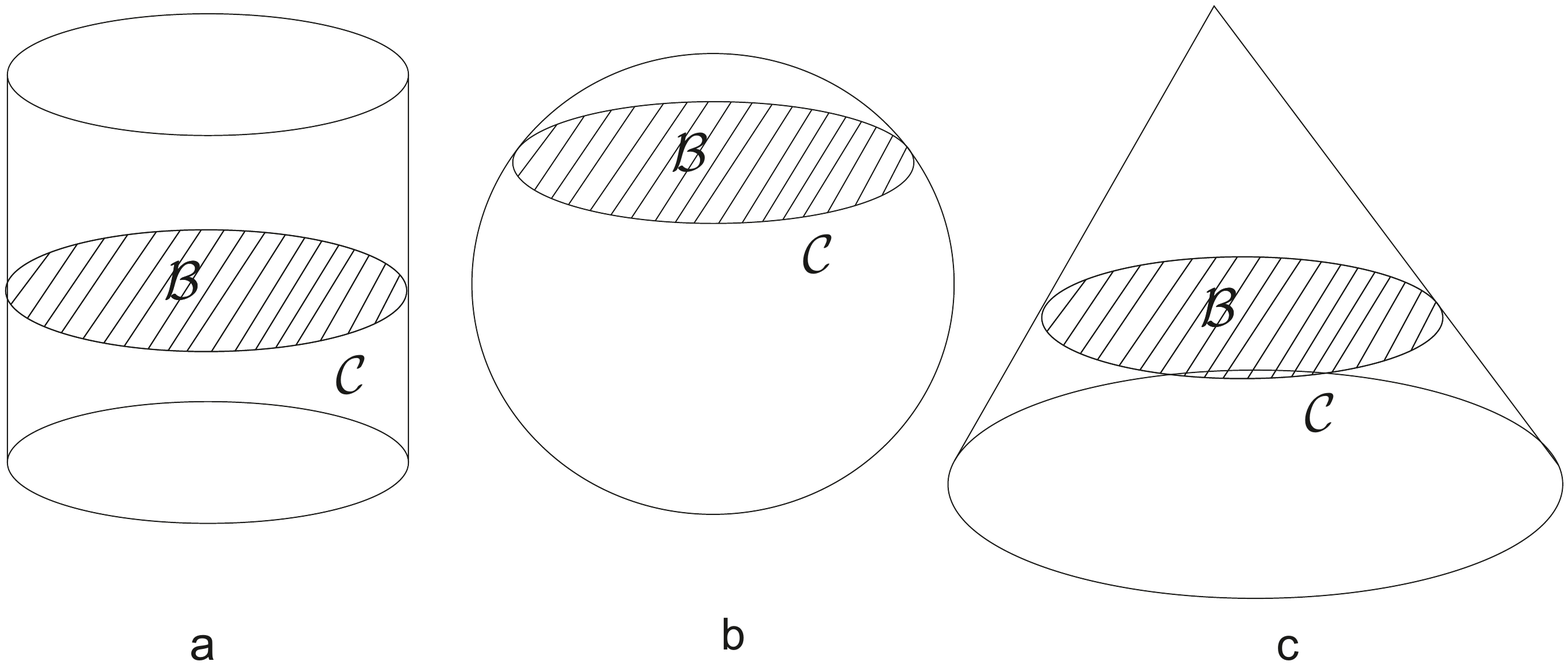}
\caption{\small{The figure shows different examples of the entangling surface
$\cal B$ crossing the boundary at $\cal C$. The boundary is a cylinder (a), sphere (b)
and a cone (c).}}
\label{F2}
\end{center}
\end{figure}

\subsection{Examples}\label{examples}

Let us discuss now some simple but not trivial examples of boundary terms in a flat
spacetime. In a flat spacetime Eqs. (\ref{3.7}), (\ref{3.10}) are simplified as
\begin{equation}\label{3.7a}
F_d={1 \over 2\pi}\int_{{\cal C}}ds~\left(3K^{(3)}_{vv}-K^{(3)}\right)~~,
\end{equation}
\begin{equation}\label{3.10a}
F_e={1 \over 2\pi}\int_{{\cal C}}ds~(N\cdot p)\left(2k_{vv}-k\right)~~,
\end{equation}
In (\ref{3.7a}) the extrinsic curvature $K_{\mu\nu}$ of
$\partial {\cal M}$ is reduced to the extrinsic curvature $K^{(3)}_{\mu\nu}$
of boundary $\partial \Sigma$ of $\Sigma$.
In (\ref{3.10a}) also
we took into account that 
one of extrinsic curvatures of $\cal B$, the one associated with a
time-like normal vector, is identically zero.
We define $k_{\mu\nu}$ as an extrinsic curvature of $\cal B$ for a space-like normal $p$
(i.e. a vector lying in a constant section $\Sigma$ and orthogonal to $\cal B$).
It is clear that $k_{\mu\nu}$ is just an extrinsic curvature of $\cal B$ in $\Sigma$.

Let us start with the invariant functional $F_d$ and consider three examples shown on 
Fig. \ref{F2}. The boundary $\partial \Sigma$ can be taken
as an infinite cylinder (a), as a sphere (b)
of the radius $R$, and as a cone
(c) with a conical angle $\varphi$. 
In all three cases the surface $\cal B$ is disc of some radius $R_B$.

For the spherical and conical cases the singular surface is tilted to the boundary.
After some simple algebra one finds with the help of (\ref{3.7a})
\begin{equation}\label{3.16a}
F_d=2~~,~~\mbox{cylindrical boundary}~~,
\end{equation}
\begin{equation}\label{3.16b}
F_d={R_B \over R}~~,~~\mbox{spherical boundary}~~,
\end{equation}
\begin{equation}\label{3.16c}
F_d=2\sqrt{1-\left({\varphi \over 2\pi}\right)^2}~~,~~\mbox{conical boundary}~~.
\end{equation}
For the cylinder $K^{(3)}=K^{(3)}_{vv}=1/R_B$. For the sphere $K^{(3)}=2K^{(3)}_{vv}=1/R$, where $R$ is the radius of the
sphere. In the  case of the cone $K^{(3)}=K^{(3)}_{vv}=\cos\alpha/R_B$, where the angle 
$\alpha$ is related to the conical angle $\varphi$ as $\varphi=2\pi \sin\alpha$.

The boundary functional $F_e$
shares a common property with the bulk functional $F_b$: the both invariants
vanish when $(k_i)_{\mu\nu}=\frac 12 h_{\mu\nu}\mbox{Tr}~ k_i$.
In particular the functionals vanish when $\cal B$ is a segment of $S^2$. $F_e$
is non-zero only when $\cal B$ is tilted to the boundary. This property differs $F_e$ from $F_d$. 

A simple example
of non-zero $F_e$ is the case of a planar boundary $\partial \Sigma$ and 
a cylindrical surface $\cal B$.  Let $\cal B$ be a cylinder of the radius $R$ and $\alpha$ be 
an angle between axis of the cylinder and normal vector $N$ to $\partial \Sigma$. One 
finds for the integrand in (\ref{3.10a})
\begin{equation}\label{3.17}
(N\cdot p)\left(2k_{vv}-k\right)={(N\cdot p) \over R} ~{\cos 2\alpha +(N\cdot p)^2
\over 1-(N\cdot p)^2}~~,
\end{equation}
where $p$ is a normal to $\cal B$ at $\cal C$.

\section{Some manifestations of boundary entanglement}\label{Phys}
\subsection{The anomalous scaling}

The quantum entanglement in the presence of physical
boundaries depends on a material of the boundary  
and on a cutoff parameter. It is possible
however to identify  
some properties of the boundary entanglement which are
cutoff independent. 

There is a simple way to get rid of the leading terms 
in entanglement entropy, Eq. (\ref{1.1}), by taking the difference 
\begin{equation}\label{4.1}
\Delta S^{(n)}\equiv S^{(n)}_a-S^{(n)}_b
\simeq \Delta s^{(n)}_{~4}~\ln(\Lambda\mu)+\Delta S^{(n)}_{\mbox{\tiny{fin}}}~~.
\end{equation}
Here $S^{(n)}_a$ and $S^{(n)}_b$ are entanglement entropies for a fixed field model but for different
geometrical configurations.
In (\ref{4.1}) the leading terms with a power dependence on the cutoff cancel out when areas of entangling surfaces as well as lengths of 
the entangling
curves for the two configurations coincide. The remaining logarithmic 
term depends on the difference $\Delta s^{(n)}_{~4}=s^{(n)}_{~4,a}-s^{(n)}_{~4,b}$. 

The last term
in the r.h.s. of (\ref{4.1}) is $\Delta S^{(n)}_{\mbox{\tiny{fin}}}=S^{(n)}_{\mbox{\tiny{fin}},a}-S^{(n)}_{\mbox{\tiny{fin}},b}$. It is related to  parts of the entropies
which are finite when the cutoff $\Lambda$ is sent to infinity.
One may call $S^{(n)}_{\mbox{\tiny{fin}}}$ a renormalized entropy.

In some simple configurations (\ref{4.1}) depends 
only on the boundary entanglement. Consider, as an example, configurations discussed in Sec. 
\ref{examples}, see Fig. \ref{F2}. This is the case of a system in a flat spacetime with a boundary 
and a flat entangling surface $\cal B$. We choose $\cal B$ to be a disc of a radius 
$R$.  The bulk terms $F_b$, $F_c$ in 
$s^{(n)}_{~4}$ equal to zero, $F_a=-2$ and cancel out as well.  For 
the given configurations  the boundary
invariant $F_e$ is absent as well. 

Suppose the system is in a ground state. Then $R$ is the single dimensional 
parameter, and one can write (\ref{4.1}) as
\begin{equation}\label{4.2}
\Delta S^{(n)}(R)=\left[d(n)(F_{d,1}-F_{d,2})+\Delta z(n)\right]\ln(\Lambda R)+
\Delta S^{(n)}_{\mbox{\tiny{fin}}}~~.
\end{equation}
When the cutoff is removed the finite part $\Delta S^{(n)}_{\mbox{\tiny{fin}}}$ 
depends only on dimensionless parameters of the system but not on $R$. 

\begin{figure}[h]
\begin{center}
\includegraphics[height=8.5cm,width=12cm]{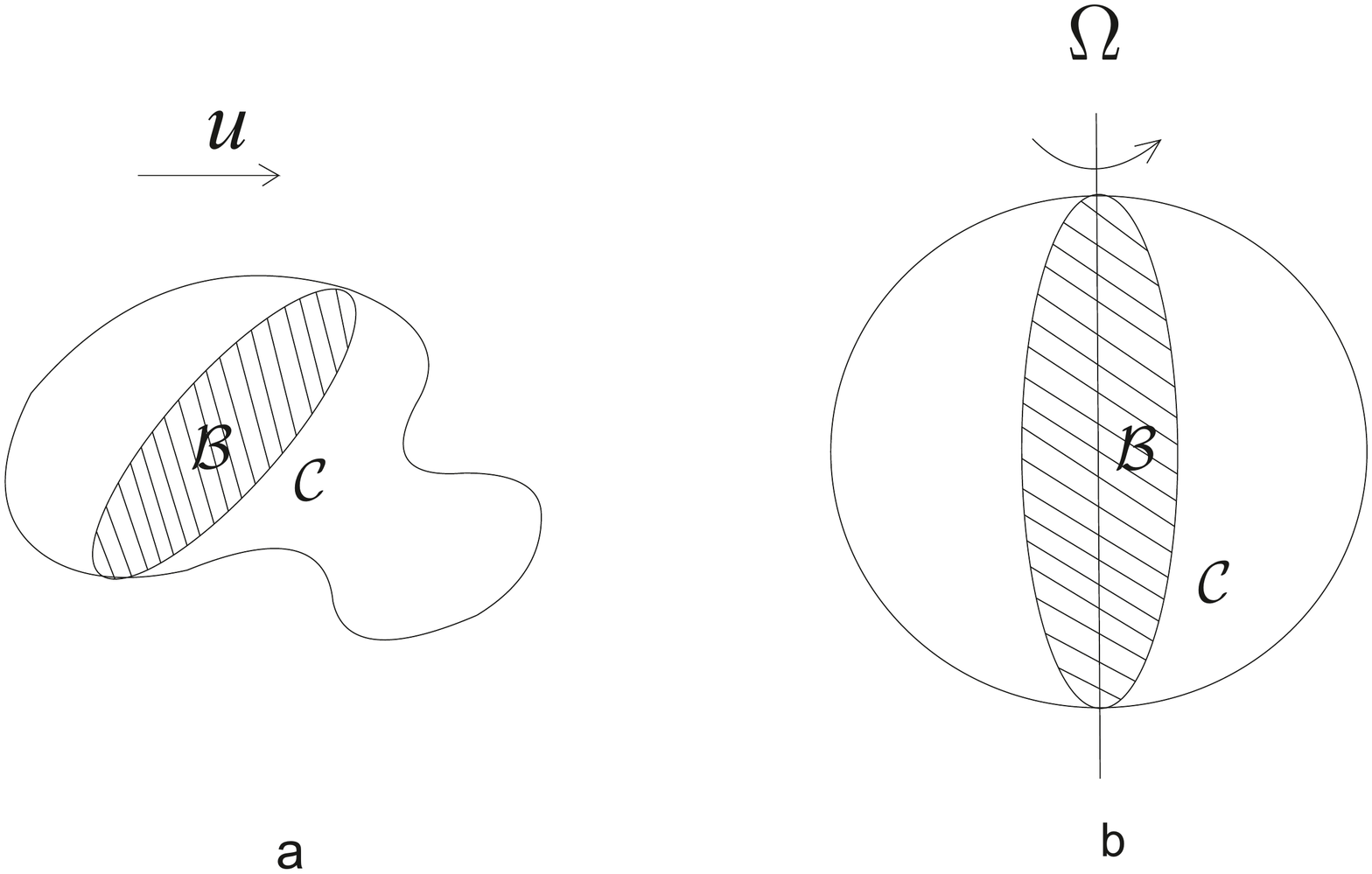}
\caption{\small{Entangling surfaces for regions
which are at rest in accelerated frames in a flat spacetime. 
Case (a) is a Rindler 
frame of reference with a velocity vector $u$, case (b) is a sphere rotating 
with a constant coordinate angular velocity $\Omega$.}}
\label{F3}
\end{center}
\end{figure}

Therefore, in the ground state one can consider the following cutoff independent
quantity:
\begin{equation}\label{4.3}
{\partial \over \partial \ln R}\Delta S^{(n)}(R)=\Delta
{\partial S^{(n)}(R) \over \partial \ln R} 
=d(n)(F_{d,1}-F_{d,2})+\Delta z(n)~~,
\end{equation}
where the derivative is taken under fixed dimensionless parameters and
$\Delta z(n)=z_1(n)-z_2(n)$. The coefficient $d(n)$ depends on the model. 
For example, for a conformal theory in a cylinder and in a cone, if zero modes are ignored,
one gets from (\ref{3.16a}), (\ref{3.16c})
\begin{equation}\label{4.4}
{\partial \over \partial \ln R}
(S^{(n)}_{\mbox{\tiny{cyl}}}(R)-S^{(n)}_{\mbox{\tiny{cone}}}(R))
=2d(n)\left(1-\sqrt{1-\left({\varphi \over 2\pi}\right)^2}\right)~~.
\end{equation}
This equation compares the evolution of the two entropies for a disc entangling surface 
under a change of the disc radius. The property that the entropies behave differently is entirely due to the boundary effects and different orientations of the disc to boundary
surfaces.

\subsection{Acceleration effects}\label{accel}

Systems which are at rest in non-inertial frames of references represent other examples where effects of the boundary entanglement in a flat spacetime 
are significant. Let us show how a non-vanishing acceleration or rotation of boundaries change the entanglement.

We start with a general case. Consider a stationary 
spacetime with a time-like Killing vector field
$\xi^\mu=\partial_\tau$. Suppose a system is at rest in a Killing frame of reference 
(a frame whose velocity four-vector $u^\mu$ is directed along $\xi^\mu$).
Physical relations in the given frame are determined
by a standard 3+1 decomposition of the metric  
\begin{equation}\label{4.5}
ds^2=B(d\tau+a_idx^i)^2+h_{ij}dx^i dx^j~~,
\end{equation} 
where $B=g_{\tau\tau}=u_0^2$, $a_i=u_i/u_0$. The 3-dimensional metric $h_{ij}$
allows one to measure physical distances between nearby observers while 
a 3-vector $a_i$ can be used to synchronize their clocks.

In 4-dimensional notations one can write $g_{\mu\nu}=h_{\mu\nu}+u_\mu u_\nu$.
Since we use a Euclidean
formulation of the theory, $u^\mu u_\mu=1$. For stationary solutions 
of the Einstein equations metric (\ref{4.5}) is obtained from the Lorentzian metric 
via a Wick rotation. For stationary but non-static metrics
this procedure also implies the Wick rotation of certain 
parameters of the solution.

The 3+1 decomposition applied to invariant (\ref{3.7}) yields the following
result:
\begin{equation}\label{4.6}
F_d=F_d^{(3)}+F_d^{(\mbox{\tiny{acc}})}~~,
\end{equation}
where $F_d^{(3)}$ is a pure 3-dimensional part, and $F_d^{(\mbox{\tiny{acc}})}$ is a part
related to acceleration and rotation of the boundary.
The first term in the r.h.s. of (\ref{4.6}) is analogous to (\ref{3.7a}) 
and is defined as
\begin{equation}\label{4.7}
F_d^{(3)}={1 \over 2\pi}\int_{{\cal C}}ds~\left(3K^{(3)}_{vv}-K^{(3)}\right)~~,
\end{equation}
where
$K^{(3)}_{\mu\nu}=h_{\mu}^\lambda h_{\nu}^\rho K_{\lambda\rho}$
and $h^\mu_\nu=\delta^\mu_\nu-u^\mu u_\nu$. As is shown in Appendix \ref{A-Killing}, 
$K^{(3)}_{\mu\nu}$ coincides with
an extrinsic curvature tensor of a boundary of a 3D space with metric $h_{ij}$, 
see Eq. (\ref{4.5}). 

Acceleration and local rotation of the boundary result in a non-trivial
term
\begin{equation}\label{4.8}
F_d^{(\mbox{\tiny{acc}})}={1 \over 2\pi}\int_{{\cal C}}ds~
\left[(w\cdot N)(1-3(v\cdot u)^2)-6(v\cdot u)\Omega_{\perp}\right]~~,
\end{equation}
where $w^\mu=\nabla_u u^\mu$ is the physical acceleration of an observer in the 
given frame, and $\Omega_{\perp}$ is a projection of the local angular velocity 3-vector 
$\Omega^i=-\frac 12 \sqrt{B}\epsilon^{ijk}a_{j,k}$ on a direction orthogonal to 
vectors $N$ and $v$. The local angular velocity describes a rotation with respect 
to a local inertial frame.  Derivation of (\ref{4.7}) and (\ref{4.8})
is presented in Appendix \ref{A-Killing}. 

Eqs. (\ref{4.7}), (\ref{4.8}) can be used to study entanglement entropies
in different non-inertial frames in a flat spacetime. 
Consider, as an example, a system which accelerates along a $z$ coordinate, see Fig. \ref{F3}.a. Suppose that it is at rest in the so called Rindler coordinates defined by metric (\ref{4.5}) with $a_i=0$,
$B=\rho^2$, $h_{ij}dx^idx^j=d\rho^2+dx^2+dy^2$, where $\rho>0$. Acceleration of
a point with a coordinate $\rho$ is $w=\sqrt{w^\mu w_\mu}=1/\rho$. 

Let us  require that the geometry of the system in the coordinates $\rho$, $x$, $y$ is the same
as in the Minkowsky spacetime. This can be required since $h_{ij}$ is a flat metric.
We also suppose that entangling surfaces $\cal B$ in the Rindler and in the
Minkwsky coordinates are identical.
Since the spacetime is flat $F_c=0$. The invariants 
$F_b$ and $F_e$ depend on extrinsic curvatures of $\cal B$, see 
(\ref{3.6}), (\ref{3.10}), and may be non-trivial. However, they coincide
with corresponding invariants for a system in Minkowsky coordinates.  The reason
is very simple: $\cal B$ is defined as a surface in a constant time slice.
Such slices in the Rindler coordinates are parts of constant time sections
in inertial frames.  Finally, since topology is fixed, $F_a$ also does not depend on 
the acceleration.

Therefore, the difference of entanglement entropies in
the Rindler and in the Minkowsky frames is determined by the $F_d$ invariants and
it is
\begin{equation}\label{4.9}
\Delta S^{(n)}=S^{(n)}_{\mbox{\tiny{Rind}}}-S^{(n)}_{\mbox{\tiny{Mink}}}=
 {d(n) \over 2\pi}\int_{{\cal C}}ds~
(w\cdot N)~\ln(\Lambda \mu)+
\Delta S^{(n)}_{\mbox{\tiny{fin}}}~~.
\end{equation}
In the accelerated frame the $F_d^{(3)}$ part coincides with the Minkowsky  invariant 
$F_{d,\mbox{\tiny{Mink}}}$. This means that $F_{d,\mbox{\tiny{Rind}}}-F_{d,\mbox{\tiny{Mink}}}=F_d^{(\mbox{\tiny{acc}})}$.
Eq. (\ref{4.9}) follows if one uses (\ref{4.8}) with $(v\cdot u)=0$. 

According to (\ref{4.9}) an acceleration in the direction orthogonal to the
boundary $\partial \Sigma$ results in a logarithmic term in the entropy.  
Acceleration along the boundary does not affect the entropy. Computations
for particular cases can be easily done,
taking into account that the only non-zero component of $w^\mu$ is $w^\rho=1/\rho$.
If the acceleration vector is strictly orthogonal to $\partial \Sigma$, $(w\cdot N)=-1$,
one finds 
from (\ref{4.9})
\begin{equation}\label{4.10}
\Delta S^{(n)}=
-{d(n) L \over 2\pi \rho}~\ln(\Lambda \mu)+
\Delta S^{(n)}_{\mbox{\tiny{fin}}}~~,
\end{equation}
where $L$ is the length of $\cal C$, $\rho$ is a coordinate of the boundary.

Another interesting case is a system  
rotating with a constant angular velocity $\Omega$ in the Minkowsky spacetime. Suppose that the system has an axial
symmetry and a symmetry axis coincides with an axis of the rotation. Let 
the entangling surface $\cal B$ lie in a plane which goes through the symmetry axis,  
as is shown on Fig. \ref{F3}.b. Since $\cal B$ is flat $F_b=F_e=0$. Also $F_c=0$, while $F_a$ does not depend on the
rotation. A computation shows
that the entanglement entropy of a rotating system differs from the entropy 
of the same system in an inertial frame as
\begin{equation}\label{4.11}
\Delta S^{(n)}=S^{(n)}_{\mbox{\tiny{Rot}}}-S^{(n)}_{\mbox{\tiny{Mink}}}=
d(n)F_d^{(\mbox{\tiny{acc}})}~\ln(\Lambda \mu)+
\Delta S^{(n)}_{\mbox{\tiny{fin}}}~~,
\end{equation}
\begin{equation}\label{4.12}
F_d^{(\mbox{\tiny{acc}})}=\frac 12\left(1-{1 \over \sqrt{1-\Omega^2R^2}}\right)~~,
\end{equation}
where $R$ is the radius of the boundary. To get this result we used (\ref{4.8}) and took into account that the velocity vector $u$ is orthogonal to $\cal B$ 
and, hence, $(u\cdot v)=0$. The acceleration vector has a single non-zero component  $w^r=\frac 12 \partial_r\ln (1-\Omega^2r^2)$, where $r$ is a distance from a point 
to the rotation axis.

\section{Discussion and open problems}\label{concl}
\setcounter{equation}0

The aim of this work was to study boundary effects in entanglement, and, in particular, 
boundary terms in anomalous scaling
of entanglement entropies for different conformal field theories 
with three spatial dimensions. Our results can describe the boundary entanglement 
of electromagnetic field excitations, as a most important physical application.

We identified 2 independent conformal structures 
in the leading boundary terms and succeeded in calculating one of the coefficient functions $d(n)$
at one of them. As was shown for simple examples,
the anomaluos boundary entanglement is sensitive to a tilt angle between an entangling surface  and the boundary.  As well it is sensitive to a curvature of the entangling curve.
Acceleration and rotation of the boundary also affect the boundary
entanglement. These leading boundary terms do not depend on the temperature of the system.

From the analysis of Sec. \ref{Phys} we conclude that boundary effects in 3 spatial dimensions not only affect the
entanglement entropies, but may define the entropies through the scaling properties,
for example, for systems with a single dimensional parameter in a ground state 
in Minkowsky
spacetime. Equation (\ref{4.3}) allows one to quantify entanglement 
of vacuum electromagnetic fluctuations. Consider the difference of entropies
$\Delta S$ for cylindrical and spherical boundaries of a radius $R$
when the entangling surfaces are flat and orthogonal to these boundaries. 
For the sphere it means that the plane lies in the equator. 
The length of the cylinder
is assumed to be infinite. A variation $\delta R$
results in the following change of the entanglement of vacuum
electromagnetic fluctuations:
\begin{equation}\label{c.14}
\delta \Delta S={1 \over 30}{\delta R \over R}~~,
\end{equation}  
where we used (\ref{3.16a}),(\ref{3.16b}),(\ref{4.3}), results from Table \ref{t1}
and ignored zero modes. Remarkably,
this change of entropies is finite, cutoff independent, and it is entirely due to
boundary effects.

We could not fix by our method the coefficient function $e(n)$. This problem is left for a future research. We have not studied non-smooth boundaries, although  boundary
defects 
yield extra contributions to the entropy  \cite{Fursaev:2006ng},\cite{Hertzberg:2010uv}. 
There are also other directions where a study of boundary effects in 3 dimensions can be continued.

One can consider conformal theories with a more general class of invariant boundary
conditions. This can be done by including background fields in the boundary
conditions in some appropriate way.
For example, for a scalar field one can impose a Robin condition of a special form
\begin{equation}\label{c.4}
\left(\partial_N+\frac 13 K\right)\varphi\mid_{\partial {\cal M}}=0~~,
\end{equation}
where the role of a background field is played by the extrinsic curvature $K$ of $\partial M$. Condition (\ref{c.4}) is invariant under conformal transformations (\ref{3.8}), (\ref{3.9}) accompanied by the corresponding transformation of the scalar field $\bar{\phi}=e^{\sigma}\phi$. The function $g(n)$ which corresponds to (\ref{c.4}) differs
from the scalar function for the Dirichlet condition by the sign. Thus, (\ref{c.4})
changes also the coefficient function $d(n)$ in the boundary entanglement. 
One can also extend these results to theories without conformal symmetry, for example, massive scalar 
and spinor fields with Dirichlet conditions,
or models where boundary conditions break conformal symmetry. 

The second interesting problem is to understand if the coefficients $d=d(1)$,
$e=e(1)$ in the entanglement entropy are new parameters of the theory or they are
related to the bulk coefficients $a$ and $c$ in the trace anomaly. The anomaly has the following
form \cite{Duff:1977ay}:
\begin{equation}\label{c.7}
\langle T_\mu^\mu\rangle=-2a~E- c~ I-{c \over 24\pi^2} ~\nabla^2 R~~,
\end{equation}
where $E$ is the volume density of the Euler characteristics, see Sec. \ref{C-notations}, and
\begin{equation}\label{c.9}
I=-{1 \over 16\pi^2}C_{\mu\nu\lambda\rho}C^{\mu\nu\lambda\rho}~~
\end{equation}
with $C_{\mu\nu\lambda\rho}$ defined in (\ref{3.5}).
In case of boundaries a more appropriate quantity is the integral of the trace anomaly with boundary terms included. The general structure of the boundary terms 
has been established in \cite{Dowker:1989ue}. The integral of $\langle T_\mu^\mu\rangle$ 
is just the heat coefficient which can be written in manifestly scale 
invariant form 
\begin{equation}\label{c.10}
\eta A_4=-a~\chi_4-c~i_4-c'j_4'-c''j_4''~~,
\end{equation}
where, as earlier, $\eta=+1$ for Bosons and $\eta=-1$ for Fermions. 
The Euler invariant $\chi_4$ is defined 
in Sec. \ref{C-notations} by Eqs. (\ref{C-1})-(\ref{C-4}),
$i_4$ is the integral of $I$, and 
\begin{equation}\label{c.11}
j'_4={1 \over 16\pi^2} \int_{\partial {\cal M}}\sqrt{H}d^3x~
\mbox{Tr}(\hat{K}^3)~~,~~j''_4={1 \over 16\pi^2} \int_{\partial {\cal M}}\sqrt{H}d^3x ~C_{\mu\nu\lambda\rho}n^\nu n^\rho\hat{K}^{\mu\lambda}~~.
\end{equation}
$\hat{K}_{\mu\nu}$ is defined in (\ref{c.12}).
The boundary terms $j_4'$ and $j_4''$ are conformal invariants since $\hat{K}_{\mu\nu}$ transforms homogeneously
under transformations (\ref{3.9}).
Coefficients $c'$, $c''$ are given in \cite{Dowker:1989ue} for a conformal scalar field with the Dirichlet condition, $c'=-2/35$, $c''=1/15$.

We note that  $a$ and $c$ in (\ref{c.7}),(\ref{c.10})  are limiting values 
at $n\to 1$ of the coefficient functions $a(n)$ and $c(n)$ introduced in (\ref{1.2}).
Eq. (\ref{3.15}) implies that $d$ is connected with the bulk coefficient $a$ as
\begin{equation}\label{c.6}
d=a+g~~.
\end{equation}
Here $g=g(1)$, $g(n)=\eta \bar{g}(n)/(1-n)$, and $\bar{g}(n)$ is defined in 
(\ref{3.13}). In certain models $d$ and $a$ may coincide,
for example, in a model which consists of two
conformal scalar fields, one with Dirichlet 
boundary condition (\ref{2.10-d}) and another with condition (\ref{c.4}).
Constants $g$ for these fields differ by the sign. Thus, for this model $d=a=2/360=1/180$.

We mentioned in Sec. \ref{intr} that the boundary $g$-function in 2D CFT's decreases under the RG flow \cite{Friedan:2003yc}.  One may expect that 
some boundary terms in four dimensions have similar features.
A possible relation between $d$ and $a$ is interesting since, as was suggested in \cite{Cardy:1988cwa}, $a$ 
might be a four dimensional analogue of the 2D Zamolodchikov  $C$-function  \cite{Zamolodchikov:1986gt}. 
There are arguments that this '$a$-function' monotonically decreases under 
a renormalization
group (RG) flow from IR to UV  fixed points, see also discussion in \cite{Komargodski:2011vj}. Similar arguments that the $a$-theorem holds for the entanglement
entropy are given in \cite{Solodukhin:2013yha}. 
An incomplete list of works where aspects of RG behaviour 
of  entanglement entropies in different dimensions have been studied 
includes \cite{Casini:2006es}-\cite{Nozaki:2012zj}. 
It would be interesting to  analyse  the boundary function $d$ along the lines of \cite{Solodukhin:2013yha}.

In general, since $d$ in (\ref{c.6}) is a sum of the bulk charge $a$ and the  seemingly independent 
constant $g$ which is determined by boundary conditions, one should not
exclude that $d$ is a new parameter of the theory. It is also worth pointing out 
a possible relation of $d$ to the bulk coefficient $c$ 
at the 'Weyl part' of the trace anomaly (\ref{c.7}).
For massless Dirac spinors one has the following expression \cite{Fursaev:2012mp}
for function $\bar{c}(n)$ defined in (\ref{3.1}):
\begin{equation}\label{c.1}
\bar{c}(n)=-{7\gamma^4+10\gamma^2-17 \over 960 \gamma}~~,
\end{equation}
with $\gamma=1/n$. (The r.h.s of (\ref{c.1}) is twice the result for the Weyl spinors
reported in \cite{Fursaev:2012mp}.)
For gauge fields \cite{Fursaev:2012mp}
\begin{equation}\label{c.2}
\bar{c}(n)={\gamma^4+10\gamma^2-11 \over 240 \gamma}~~,
\end{equation}
By comparing (\ref{c.1}), (\ref{c.2}) with results collected in Table \ref{t1} 
we find a set of 'magic' identities between the boundary and bulk coefficients
\begin{equation}\label{c.3}
\bar{d}(n)={\bar{c}(n) \over 3}~~,~~d(n)={c(n) \over 3}~~~,
\end{equation}
which hold for all arguments $n$ for the gauge field and spin 1/2 massless Dirac field. 

Formulas (\ref{c.3}) are not universal and do not apply in case of 
a conformal scalar field where $\bar{d}(n)$ cannot be represented as a linear
combination of the corresponding scalar functions $\bar{a}(n)$ and $\bar{c}(n)$.
One may point out, however, another 'magic' relation 
\begin{equation}\label{c.5}
\bar{d}_{\mbox{\tiny{scalar}}}(n)=\frac 12 \bar{d}_{\mbox{\tiny{gauge}}}(n)~~~,
\end{equation} 
which holds between scalar and gauge functions, see Table \ref{t1}. Interpretation
of (\ref{c.5}) is that each degree of freedom of electromagnetic field has the same 
entanglement as a scalar quantum.

Finally, another research direction is in studying a holographic representation
of our results for the boundary entanglement. The four dimensional conformal 
theory which admits a dual description in terms of the AdS gravity one dimension higher 
is the $SU(N)$ supersymmetric Yang-Mills theory.
Our results applied to a multiplet of fields in this model in the limit 
of a weak coupling and large $N$ yield the value $d=N^2/10$ for the total boundary charge
(for the Weyl spinors we just take half of the coefficient of Dirac spinors).
It is interesting to see if Eqs. (\ref{1.1}),(\ref{1.2}) with this coefficient
can be reproduced by applying the holographic
formula of \cite{Ryu:2006bv}.
We hope this can be done by taking into account a recent progress 
\cite{Takayanagi:2011zk} in understanding the holographic
formulation of CFT's with boundaries.

\bigskip
\bigskip
\bigskip

\noindent
{\bf Acknowledgement}

\bigskip

The author is grateful to I.L. Buchbinder, G. Esposito, A. Yu. Kamenshchik, D.V. Vassilevich  
for helpful discussions. This work was supported by RFBR grant 13-02-00950.

\newpage
\appendix

\section{Main notations and definitions}\label{C-notations}
\setcounter{equation}0

For a convenience we collect here main geometrical notations used in the paper.

\bigskip

$\cal M$: a four-dimensional manifold  (with the Euclidean signature);

$\partial {\cal M}$: a boundary of $\cal M$;

$N$: a unit outward pointing
normal vector to $\partial {\cal M}$;

$\nabla_N=N^\mu \nabla_\mu$: a normal derivative operator;

$K_{\mu\nu}=H_{\mu}^\lambda H_{\nu}^\rho N_{\lambda;\rho}$: an 
extrinsic curvature tensor of $\partial {\cal M}$, $K=g^{\mu\nu}K_{\mu\nu}$;

$H_{\mu}^\nu=\delta_{\mu}^\nu-N_{\mu}N^\nu$: a projector on a tangent space of 
$\partial {\cal M}$;

$\tau$: a time coordinate of $\cal M$;

$\xi^\mu\partial_\mu=\partial_\tau$: a Killing vector field of $\cal M$, $\xi^\mu$
is time-like in the Lorenzian signature;

$u^\mu$: a unit velocity four-vector of the Killing frame, $u^\mu$ is directed along $\xi^\mu$;

$w^\mu=\nabla_u u^\mu$ is an acceleration of the Killing frame;

$K^{(3)}_{\mu\nu}=h_{\mu}^\lambda h_{\nu}^\rho K_{\lambda\rho}$,
where $h^\mu_\nu=\delta^\mu_\nu-u^\mu u_\nu$, is a 3-dimensional 
extrinsic curvature in the Killing frame;

$\Sigma$: a constant $\tau$ hypersurface;

$\partial \Sigma$: a boundary of $\Sigma$;

$\cal B$: a co-dimension 2 entangling surface which lies in $\Sigma$;

$p_i$: a pair of ortho-normalized normal vectors to $\cal B$;

$(k_i)_{\mu\nu}=P_{\mu}^\lambda P_{\nu}^\rho (p_i)_{\lambda;\rho}$: 
extrinsic curvatures of $\cal B$, $k_i=g^{\mu\nu}(k_i)_{\mu\nu}$;
 
$P_{\mu}^\lambda=\delta_{\mu}^\lambda-(p_i)_{\mu}(p_i)^\lambda$: 
a projector on a tangent space of $\cal B$;

${\cal C}={\cal B} \cap\partial \Sigma$: an entangling curve;

$l$: a unit outward directed vector 
in a tangent space to $\cal B$, $l$ is orthogonal to $\cal C$;

$v$: a unit tangent vector to $\cal C$; 

$a=\nabla_v v$: an acceleration of $\cal C$;

$m_a$: an ortho-normalized pair of vectors at $\cal C$ which are orthogonal
to $v$ and $N$;

$K_{vv}=K_{\mu\nu}v^\mu v^\nu$, $(k_i)_{vv}=(k_i)_{\mu\nu}v^\mu v^\nu$: 
quantities defined  on $\cal C$.

The definition of the Euler characteristic $\chi_4$ on a four-dimensional manifold 
$\cal M$ with the boundary $\partial {\cal M}$ is \cite{Dowker:1989ue}
\begin{equation}\label{C-1}
\chi_4=B_4[{\cal M}]+S_4[\partial {\cal M}]~~,
\end{equation}
\begin{equation}\label{C-2}
B_4[{\cal M}]=\int_{\cal M}\sqrt{g}d^4x~E={1 \over 32\pi^2}\int_{\cal M}\sqrt{g}d^4x\left[R^2-4R_{\mu\nu}R^{\mu\nu}+
R_{\mu\nu\lambda\rho}R^{\mu\nu\lambda\rho}\right]~~,
\end{equation}
\begin{equation}\label{C-3}
S_4[\partial {\cal M}]=-{1 \over 4\pi^2}\int_{\partial {\cal M}}\sqrt{H}d^3x\left[\det K_{\mu\nu}+\hat{G}^{\mu\nu}K_{\mu\nu}\right]~~,
\end{equation}
\begin{equation}\label{C-4}
\hat{G}^{\mu\nu}=\hat{R}^{\mu\nu}-\frac 12H^{\mu\nu} \hat{R}~~,
\end{equation}
where $\hat{G}^{\mu\nu}$ is the Einstein tensor on $\partial {\cal M}$. 
The use of $H_{\mu\nu}$ in (\ref{C-4}) is equivalent to use of to induced metric
tensor.

\section{Heat coefficients for entangling surfaces 
orthogonal to boundaries}\label{A-HKC}
\setcounter{equation}0

Let $\cal M$ be a domain with a boundary $\partial {\cal M}$ in $R^4$. In a constant time
section $\Sigma$ of $\cal M$ we choose a $z$ coordinate along 
one of $R^1$ directions
and assume that $\Sigma={\cal B}\times R^1$, where $\cal B$ is a constant $z$ 
section of $\Sigma$. Thus, $\cal B$ is flat.
The boundary of $\Sigma$ is $\partial \Sigma ={\cal C} \times R^1$, where ${\cal C}=\partial {\cal B}$. 
If ${\cal C}$ is closed $\Sigma$ has a topology of a cylinder. 
Hence, ${\cal M}={\cal B}\times R^2$ and
the spacetime boundary is $\partial {\cal M}={\cal C} \times R^2$.

Construction of replicated versions ${\cal M}_n$ of $\cal M$ leads to manifolds which have 
the structure ${\cal M}_n={\cal B}\times {\cal M}_n^{(2)}$.
Here ${\cal M}_n^{(2)}$ is a 2D manifold with conical singularities which has been explicitly described in \cite{Fursaev:2006ng}. We calculate now contributions to the heat coefficient
$A_4$ on ${\cal M}_n$ for operators and boundary conditions
discussed in Sec. \ref{BC}.

\bigskip

{\bf Scalar Laplacian}. The heat kernel $K_n(\Delta;t)$ 
on ${\cal M}_n={\cal B}\times {\cal M}_n^{(2)}$ can be written as
\begin{equation}\label{A1.1}
K_n(\Delta;t)=K_n(\Delta_{(2)},t)K(\Delta_{\cal B};t)~~,
\end{equation}
where $K_n(\Delta_{(2)},t)$ and $K(\Delta_{\cal B};t)$
are heat kernels on  ${\cal M}_n^{(2)}$  and $\cal B$, respectively.
Consider asymptotics
\begin{equation}\label{A1.2}
K_n(\Delta_{(2)};t)\simeq  \sum^\infty_{p=0} A_p(\Delta_{(2)},n)
~t^{(p-2)/2}~~,~~t\to 0~~,
\end{equation}
\begin{equation}\label{A1.3}
K(\Delta_{\cal B};t)\simeq  \sum^\infty_{p=0} A_p(\Delta_{\cal B})
~t^{(p-2)/2}~~,~~t\to 0~~.
\end{equation}
We are interested in the singular part $\bar{A}_4(n)$ of $A_4$, see (\ref{3.1}),
which comes from conical singularities. It follows from (\ref{A1.1})-(\ref{A1.3}) that
\begin{equation}\label{A1.4}
\bar{A}_4(n)=\bar{A}_2(n)A_2(\Delta_{\cal B})~~,
\end{equation}
where $\bar{A}_2(n)$ is the singular part of $A_2(\Delta_{(2)},n)$.
It can be shown that
\begin{equation}\label{A1.6}
\bar{A}_2(n)={1 \over 12\gamma}\left(\gamma^2 -1\right)~~,
\end{equation}
where $\gamma=1/n$, and for the Dirichlet boundary condition
\begin{equation}\label{A1.7}
A_2(\Delta_{\cal B})={1 \over 12\pi}\int_{\cal C}k_B ds~~.
\end{equation}
(Eq. (\ref{A1.7}) also holds for the Neumann condition, which we use for the ghost 
operator.)
Thus, $\bar{A}_4(n)$ is a purely boundary term. Substitution
of (\ref{A1.6}),(\ref{A1.7}) in (\ref{A1.4}) yields formula (\ref{3.13}), where
the factor $\bar{g}(n)$ for the scalar Laplacian is given in Table \ref{t1}.

\bigskip

{\bf Spinor Laplacian}.  We take the spinor Laplacian and use the boundary 
conditions discussed in Sec. \ref{BC}. Eqs. (\ref{A1.1}), (\ref{A1.4})  hold 
for the spinor operators as well but expressions (\ref{A1.6}),(\ref{A1.7}) 
for the heat coefficients of 2D spinor operators are modified.
\begin{equation}\label{A1.8}
\bar{A}_2(n)=-{1 \over 12\gamma}\left(\gamma^2 -1\right)~~,
\end{equation}
\begin{equation}\label{A1.9}
A_2(\Delta_{\cal B})=-{1 \over 12\pi}\int_{\cal C}k_B ds~~.
\end{equation}
The above spinor coefficients differ from scalar coefficients (\ref{A1.6}),(\ref{A1.7}) only by
the sign.
Functions $g(n)$, $\bar{d}(n)$ and $d(n)$ for the spinor coefficient $A_4(\Delta_n)$,
which are listed in Table \ref{t1}, follow from (\ref{A1.2}),(\ref{A1.3}). 

Derivation of (\ref{A1.8}) can be found, for example, in \cite{Fursaev:1996uz}.
Several comments are in order concerning (\ref{A1.9}).  A self-consistent 
formulation of boundary problem (\ref{2.10-dd}) for the spinor 
Laplacian requires  additional condition
\begin{equation}\label{A1.10}
(\nabla_N-S)\Pi_+ \psi\mid_{\partial {\cal M}}=0~~,
\end{equation}
where $\Pi_+=1-\Pi_-$, $S=-\frac 12 K\Pi_+$, see details in \cite{Vassilevich:2003xt}. 
(For the considered case $K=k_B$.)
Thus, spinor boundary conditions are the mixed ones.
For the two dimensional spin 1/2 Laplacian $\Delta_{\cal B}$
acting on 2D spinors $\psi_2$ the corresponding boundary condition is
\begin{equation}\label{A1.10-b}
(\nabla_N-S_2)(\Pi_+ )_2\psi_2\mid_{\cal C}=0~~,
\end{equation}
with $S_2=-\frac 12 k_B(\Pi_+)_2$ and a 2D projector $(\Pi_+)_2$.
The boundary part of $A_2$ in this case is
\begin{equation}\label{A1.11}
A_2(\Delta_{\cal B})={1 \over 12\pi}\int_{\cal C} ds~\mbox{Tr}_2(k_B+6S_2)~~.
\end{equation}
To get (\ref{A1.9}) from (\ref{A1.11}) one should take into account that 
$\mbox{Tr}_2I=2$, $\mbox{Tr}_2(\Pi_+)_2=1$.

\bigskip

{\bf Coefficients for gauge fields}.
We need to calculate contributions  from boundary conical singularities
to 'total' heat coefficient (\ref{2.11}) which is
\begin{equation}\label{A.12}
\bar{A}_4^{(\mbox{\tiny gauge})}=\bar{A}_4(\Delta)-2
\bar{A}_4(\Delta^{(\mbox{\tiny gh})})~~,
\end{equation}
where $\Delta=\Delta^{(1)}$ is the vector Laplacian defined in Sec. \ref{BC}.
The ghost operator $\Delta^{(\mbox{\tiny gh})}$ is just the scalar Laplacian
with minimal coupling and Neumann boundary condition (\ref{2.10-gg}). As a result, 
$\bar{A}_4(\Delta^{(\mbox{\tiny gh})})$ coincides with the scalar coefficient.

Calculation of the vector coefficient is more tricky.
Let us denote coordinates on ${\cal B}$ and ${\cal M}_n^{(2)}$
as $x$ and $y$, respectively. The vector field on 
${\cal M}_n={\cal B}\times {\cal M}_n^{(2)}$ is decomposed as
\begin{equation}\label{A1.16}
V^\mu(x,y)=b^\mu(x)\varphi(y)+d^\mu(y)\chi(x)~~,
\end{equation} 
where vector $b^\mu(x)$ is tangent to ${\cal B}$ and
$d^\mu(y)$ is tangent to ${\cal M}_n^{(2)}$. Vectors $b^\mu(x)$, $d^\mu(y)$
have two non-vanishing components each,  $\varphi(y)$, $\chi(x)$ are
scalars.

Instead of (\ref{A1.1}) for the vector Laplacian $\Delta^{(1)}$ one has
\begin{equation}\label{A1.13}
K_n(\Delta;t)=K_{n}(\Delta_{(2)},t)K_{\parallel}(\Delta_{\cal B};t)+
K_{n,\parallel}(\Delta_{(2)},t) K(\Delta_{\cal B};t)~~.
\end{equation}
The first term in the r.h.s. of (\ref{A1.13}) corresponds to the component
tangent to ${\cal B}$. Here $K_{\parallel}(\Delta_{\cal B};t)$ 
is the heat kernel of 2D vector Laplacian for $b^\mu(x)$, 
$K_n(\Delta_{(2)},t)$ is the scalar heat kernel for $\varphi(y)$.
The second term in the r.h.s. of (\ref{A1.13}) is for the component tangent to 
${\cal M}_n^{(2)}$. Therefore, $K_{n,\parallel}(\Delta_{(2)},t)$ is 
the heat kernel of 2D vector Laplacian for  $d^\mu(y)$ on ${\cal M}_n^{(2)}$, 
$K(\Delta_{\cal B};t)$ is the scalar heat kernel for $\chi(x)$.

Eq. (\ref{A1.4}) for the heat coefficient in case of $\Delta^{(1)}$ is replaced with
\begin{equation}\label{A1.14}
\bar{A}_4(n)=\bar{A}_2(n)A_{2,\parallel}(\Delta_{\cal B})+
\bar{A}_{2,\parallel}(n)A_2(\Delta_{\cal B})~~,
\end{equation}
where $\bar{A}_2(n)$, $A_{2,\parallel}(\Delta_{\cal B})$, 
$\bar{A}_{2,\parallel}(n)$, $A_2(\Delta_{\cal B})$ correspond to
$K_{n}(\Delta_{(2)},t)$, $K_{\parallel}(\Delta_{\cal B};t)$,
$K_{n,\parallel}(\Delta_{(2)},t)$, $K(\Delta_{\cal B};t)$, respectively.

$\bar{A}_2(n)$ and $\bar{A}_{2,\parallel}(n)$ are singular parts 
of coefficients of  scalar and vector Laplacians on 2D manifolds with conical singularities
($d^\mu(y)$ and $\varphi(y)$ fields).
$\bar{A}_2(n)$ is given by (\ref{A1.6}), and 
\begin{equation}\label{A1.15}
\bar{A}_{2,\parallel}(n)={1 \over 6\gamma}\left(\gamma^2 -1\right)+{1 \over \gamma}-1
={1 \over 6\gamma}\left(\gamma^2 -6\gamma+5\right)~~,
\end{equation}
see, for example, \cite{Fursaev:1996uz}.

To calculate  $A_2(\Delta_{\cal B})$, $A_{2,\parallel}(\Delta_{\cal B})$ one has to use 
boundary conditions (\ref{2.10-gv}). The normal vector $N$ to
$\partial{\cal M}$ is tangent to $\cal B$, $K_{\mu\nu}$ does not have components orthogonal to $\cal B$. This yields the following conditions:
\begin{equation}\label{A1.17}
\partial_N\chi\mid_{\cal C}=0~~,~~(b\cdot N)\mid_{\cal C}=0~~,~~
(\partial_N b^i-S^i_jb^j)\mid_{\cal C}=0~~,
\end{equation}
where $S^i_j=-(k_B)^i_j$, and $b^i$ are the components
of $b^\mu$ tangent to $\cal C$. $A_2(\Delta_{\cal B})$ is given by (\ref{A1.7}).
Boundary conditions for the vector fields are the mixed conditions. One has 
\begin{equation}\label{A1.18}
A_{2,\parallel}(\Delta_{\cal B})={1 \over 12\pi}\int_{\cal C} ds~\mbox{Tr}_{\parallel}(k_B+6S)=
-{1 \over 3\pi}\int_{\cal C} ds~k_B~~,
\end{equation}
where $\mbox{Tr}_{\parallel} I=2$, $\mbox{Tr}_{\parallel} S=-k_B$.
By collecting all results above and using (\ref{A.12}) one gets
\begin{equation}\label{A1.19}
\bar{A}_4(\Delta)=-{1 \over 72\pi\gamma}(\gamma^2+6\gamma-7)\int_{\cal C} ds~k_B~~,
\end{equation}
\begin{equation}\label{A1.20}
\bar{A}_4^{(\mbox{\tiny gauge})}=-{1 \over 36\pi\gamma}(\gamma^2+3\gamma-4)\int_{\cal C} ds~k_B~~.
\end{equation}
Eq. (\ref{A1.20}) yields necessary functions $g(n)$, $\bar{d}(n)$ and $d(n)$
listed in Table \ref{t1}.

\section{Boundary effects in the Killing frame}\label{A-Killing}
\setcounter{equation}0

Here we give a more detailed description of results in the Killing frame 
discussed in Sec. \ref{accel}. The first covariant derivative
of the velocity 4-vector allows the following presentation:
\begin{equation}\label{B.1}
u_{\mu;\nu}=w_\mu u_\nu+\Omega_{\mu\nu}~~,
\end{equation}
\begin{equation}\label{B.2}
\Omega_{\mu\nu}=\frac 12 h_\mu^\lambda h_\nu^\rho (u_{\lambda;\rho}-u_{\rho;\lambda})~~.
\end{equation}
Quantity (\ref{B.2}) is related to an antisymmetric part of $u_{\mu;\nu}$ and is
called a rotation tensor.
A projection of the symmetric part of $u_{\mu;\nu}$ (called a deformation tensor)
vanishes in the Killing frame.

We are interested in 3+1 decomposition of boundary invariants. One can easily see
that the normal vector to $\partial {\cal M}$ is orthogonal to 4-velocity vector,
$(N\cdot u)=0$. The tangent vector $v$ to the entangling curve may be not orthogonal 
to $u$ in stationary but non-static spacetimes (when $u$ is not orthogonal to constant
time sections). Therefore, we should keep $(v\cdot u)$ non-vanishing in all expressions.

We define a 3-dimensional projection of the boundary extrinsic curvature
tensor onto directions orthogonal to the velocity 4-vector
\begin{equation}\label{B.3}
K^{(3)}_{\mu\nu}=h_{\mu}^\lambda
h_{\nu}^\rho K_{\lambda\rho}=h_{\mu}^\lambda H_\lambda^\alpha h_{\nu}^\rho 
H_\rho^\beta N_{\alpha;\beta}~~,
\end{equation}
where $H_\lambda^\alpha=\delta_\lambda^\alpha-N_\lambda N^\alpha$.
Since $(N\cdot u)=0$ one can write  (\ref{B.3}) also as
\begin{equation}\label{B.4}
K^{(3)}_{\mu\nu}=H_{\mu}^\lambda
H_{\nu}^\rho N_{\lambda \mid \rho}~~,~~~
N_{\lambda \mid \rho}=h_\lambda^\alpha h_\rho^\beta N_{\alpha;\beta}~~.
\end{equation}
This shows that $K^{(3)}_{\mu\nu}$ is an extrinsic curvature of the boundary on
3D space with metric $h_{ij}$.  The trace of $K_{\mu\nu}$ can be written as
\begin{equation}\label{B.5}
K=K^{(3)}+K_{uu}=K^{(3)}-(w \cdot N)~~,
\end{equation}
where $K^{(3)}$ is the trace of $K^{(3)}_{\mu\nu}$, and 
$K_{uu}=u^\mu u^\nu K_{\mu\nu}=u^\mu u^\nu N_{\mu;\nu}=-(w \cdot N)$.

The tangent vector $v$ to $\cal C$ can be decomposed as 
$v=v_{\perp}+(v\cdot u) u$, where $v_{\perp}$ is a component of $v$ orthogonal to 4-velocity vector, $(v_{\perp}\cdot u)=0$. Therefore, 
on $\cal C$
\begin{equation}\label{B.6}
K_{vv}=K^{(3)}_{vv}+(v \cdot u)^2 K_{uu}+ 2(v \cdot u)
u^\mu v^\nu_{\perp}K_{\mu\nu}~~,
\end{equation}
where $K^{(3)}_{vv}=v^\mu v^\nu K^{(3)}_{\mu\nu}$ is a component of 3D extrinsic curvature
along $v$. We can use now (\ref{B.2}) to make further transformations 
\begin{equation}\label{B.7}
u^\mu v^\nu_{\perp}K_{\mu\nu}=u^\mu v^\nu_{\perp}N_{\mu;\nu}=-
N^\mu v^\nu_{\perp}u_{\mu;\nu}=-\Omega_{\perp}~~,
\end{equation}
where $\Omega_{\perp}=N^\mu v^\nu_{\perp}\Omega_{\mu\nu}=N^i v^i\Omega_{ij}$.
One can show that $\Omega_{ij}=-\frac 12 \sqrt{B}(a_{i,j}-a_{j,i})$, where
we used (\ref{4.5}). In 3D notations $\Omega_{\perp}=
(\overrightarrow{\Omega} \cdot [\overrightarrow{N}\times \overrightarrow{v}])$, where $(\overrightarrow{\Omega})_i=\frac 12
\epsilon_{ijk}\Omega^{jk}$. Eq. (\ref{B.6}) takes the form
\begin{equation}\label{B.8}
K_{vv}=K^{(3)}_{vv}-(v \cdot u)^2(w \cdot N) - 2(v \cdot u)\Omega_{\perp}~~.
\end{equation}
As a result we get
\begin{equation}\label{B.9}
3K_{vv}-K=(3K^{(3)}_{vv}-K^{(3)})+(w \cdot N)(1-3(v \cdot u)^2) - 6(v \cdot u)\Omega_{\perp}~~.
\end{equation}
This yields relations (\ref{4.6})-(\ref{4.8}).


\newpage

\begin{thebibliography}{}

\bibitem{Fursaev:2006ng} D.V. Fursaev, {\it Entanglement Entropy in Critical Phenomena and Analogue Models of Quantum Gravity},
Phys. Rev. {\bf D73} (2006) 124025
e-Print: hep-th/0602134.


\bibitem{Hertzberg:2010uv} M.P. Hertzberg, F. Wilczek, {\it Some Calculable 
Contributions to Entanglement Entropy}, Phys. Rev. Lett. {\bf 106} (2011) 050404,
e-Print: arXiv:1007.0993.

\bibitem{Fursaev:2012mp} D.V. Fursaev, JHEP 1205 (2012) 080,
{\it Entanglement Renyi Entropies in Conformal Field Theories and Holography}, 
e-Print: arXiv:1201.1702 [hep-th].


\bibitem{Calabrese:2009qy} J.L. Cardy, P. Calabrese, {\it Entanglement Entropy and Conformal Field Theory}, J. Phys. {\bf A42} (2009) 504005,
e-Print: arXiv:0905.4013 [cond-mat.stat-mech]. 

\bibitem{Cardy:2004hm}  J.L. Cardy, {\it Boundary Conformal Field Theory}, 
e-Print: hep-th/0411189.

\bibitem{Affleck:1991tk} I. Affleck and A. Ludwig, {\it Universal Noninteger 'Ground State Degeneracy' in Critical Quantum Systems}, Phys. Rev. Lett. {\bf 67} (1991) 161-164.

\bibitem{Friedan:2003yc} D. Friedan andf A. Konechny, {\it On the Boundary Entropy of One-dimensional Quantum Systems at Low Temperature}, Phys. Rev. Lett. {\bf 93} (2004) 030402,
e-Print: hep-th/0312197.

\bibitem{Fursaev:1996uz} D.V. Fursaev and  G. Miele, 
{\it Cones, Spins and Heat Kernels}, Nucl. Phys. {\bf B484} (1997) 697,
e-Print: hep-th/9605153.

\bibitem{Vassilevich:2003xt} D.V. Vassilevich, {\it Heat Kernel Expansion: User's Manual},  Phys. Rept.  {\bf 388} (2003) 279-360, e-Print: hep-th/0306138. 


\bibitem{Dowker:1989gw} J.S. Dowker, {\it Conformal Properties Of The Heat-kernel Expansion: Application To The Effective Lagrangian}, Phys. Rev. {\bf D39} (1989) 1235.

\bibitem{Fursaev:2011zz} D. Fursaev and D. Vassilevich, {\it Operators, Geometry and Quanta:
Methods of Spectral Geometry in Quantum Field Theory}, Springer Series 'Theoretical and Mathematical Physics',
Springer, 2011.

\bibitem{Buchbinder:1984} I.L. Buchbinder, {\it Renormalization Group Equations In Curved Space-time},
Theor. Math. Phys. {\bf 61} (1984) 1215.

   
\bibitem{Duff:1977ay} M.J. Duff, {\it Observations on Conformal Anomalies}, Nucl.
Phys. {\bf B125} (1977) 334. 
  
\bibitem{Dowker:1989ue}  J.S. Dowker, J.P. Schofield, {\it Conformal Transformations And The Effective Action In The Presence Of Boundaries},  J. Math. Phys. {\bf 31} (1990) 808.

\bibitem{Cardy:1988cwa} J.L. Cardy, {\it Is There a C-theorem in Four-Dimensions?}, Phys. Lett. {\bf B215} (1988) 749.
  
\bibitem{Zamolodchikov:1986gt} A.B. Zamolodchikov, {\it Irreversibility of the Flux of the Renormalization Group in a 2D Field Theory},  JETP Lett. {\bf 43} (1986) 730.


\bibitem{Komargodski:2011vj}  Z. Komargodski, A. Schwimmer, 
{\it On Renormalization Group Flows in Four Dimensions}, 
JHEP {\bf 1112} (2011) 099, e-Print: arXiv:1107.3987 [hep-th].  

\bibitem{Solodukhin:2013yha} S.N. Solodukhin,{\it The a-theorem and Entanglement Entropy},
e-Print: arXiv:1304.4411 [hep-th].

\bibitem{Casini:2006es} H. Casini, M. Huerta, {\it A c-theorem for the Entanglement Entropy}, J. Phys. {\bf A40} (2007) 7031-7036, e-Print: cond-mat/0610375. 


\bibitem{Myers:2010tj} R.C. Myers and A.Sinha,
{\it Holographic c-theorems in Arbitrary Dimensions}, JHEP {\bf 1101} (2011) 125,
e-Print: arXiv:1011.5819 [hep-th].

\bibitem{deBoer:2011wk} J. de Boer, M. Kulaxizi, A. Parnachev, 
{\it Holographic Entanglement Entropy in Lovelock Gravities},
JHEP {\bf 1107} (2011) 109, e-Print: arXiv:1101.5781 [hep-th]

\bibitem{Klebanov:2012va} I.R. Klebanov, T. Nishioka, S.S. Pufu, B.R. Safdi,
{\it Is Renormalized Entanglement Entropy Stationary at RG Fixed Points?}, JHEP 
{\bf 1210} (2012) 058, e-Print: arXiv:1207.3360 [hep-th].

\bibitem{Casini:2012ei} H. Casini, M. Huerta, {\it On the RG Running of the Entanglement Entropy of a Circle}, Phys. Rev. {\bf D85} (2012) 125016,
e-Print: arXiv:1202.5650 [hep-th]. 

\bibitem{Nozaki:2012zj} M. Nozaki,  S. Ryu, T. Takayanagi,
{\it Holographic Geometry of Entanglement Renormalization in Quantum Field Theories}, 
JHEP {\bf 1210} (2012) 193,
e-Print: arXiv:1208.3469 [hep-th]. 

\bibitem{Ryu:2006bv} S. Ryu and T. Takayanagi, 
{\it Holographic Derivation of Entanglement Entropy from AdS/CFT}, 
Phys. Rev. Lett. {\bf 96} (2006) 181602,
e-Print: hep-th/0603001.

\bibitem{Takayanagi:2011zk} T. Takayanagi, {\it Holographic Dual of BCFT},
Phys. Rev. Lett. {\bf 107} (2011) 101602,
e-Print: arXiv:1105.5165 [hep-th].





\end{thebibliography}
\end{document}